\documentclass[useAMS,usenatbib,fleqn]{mn2e}
\pdfoutput=1
\usepackage{amsmath}
\usepackage{amssymb}
\usepackage{graphicx}
\usepackage{tabularx}
\usepackage{booktabs}
\usepackage{multirow}
\usepackage{array}
\usepackage{caption}
\usepackage{siunitx}
\usepackage{float}
\usepackage{placeins}
\usepackage{color}

\def\lesssim{\la}
\def\gtrsim{\ga}

\title[Towards a more realistic sink particle algorithm]{Towards a more realistic sink particle algorithm for the RAMSES code}
\author[Andreas Bleuler, Romain Teyssier]{
Andreas Bleuler, Romain Teyssier \\
{Institute for Computational Science, University of Zurich, Zurich, Switzerland}\\
{ableuler@physik.uzh.ch}  \\
}

\newcommand{\Msol}{\mathrm M_{\odot} }
\newcolumntype{L}[1]{>{\raggedright\let\newline\\\arraybackslash\hspace{0pt}}m{#1}}
\newcolumntype{C}[1]{>{\centering\let\newline\\\arraybackslash\hspace{0pt}}m{#1}}
\newcolumntype{R}[1]{>{\raggedleft\let\newline\\\arraybackslash\hspace{0pt}}m{#1}}
\DeclareSIUnit\Msun{\ensuremath{\mathrm M_{\odot}}}
\DeclareSIUnit\cm{\ensuremath{\mathrm{cm}}}
\DeclareSIUnit\year{\ensuremath{\mathrm{yr}}}
\DeclareSIUnit\parsec{\ensuremath{\mathrm{pc}}}

\begin{document}

%\date{Accepted Year Month Day. Received Year Month Day; in original form Year Month Day}
\pagerange{\pageref{firstpage}--\pageref{lastpage}} \pubyear{2014}
\label{firstpage}
\maketitle

\begin{abstract}
We present a new sink particle algorithm developed for the Adaptive Mesh Refinement code \textsc{ramses}. Our main addition is the use of a clump finder to identify density peaks and their associated regions (the peak patches). This allows us to unambiguously define a discrete set of dense molecular cores as potential sites for sink particle formation. Furthermore, we develop a new scheme to decide if the gas in which a sink could potentially form, is indeed gravitationally bound and rapidly collapsing. This is achieved using a general integral form of the virial theorem, where we use the curvature in the gravitational potential to correctly account for the background potential. We detail all the necessary steps to follow the evolution of sink particles in turbulent molecular cloud simulations, such as sink production, their trajectory integration, sink merging and finally the gas accretion rate onto an existing sink. We compare our new recipe for sink formation to other popular implementations. Statistical properties such as the sink mass function, the average sink mass and the sink multiplicity function are used to evaluate the impact that our new scheme has on accurately predicting fundamental quantities such as the stellar initial mass function or the stellar multiplicity function.
\end{abstract}


%%%%%%%%%%%%%%%%%%%%%%%%
%%%%%%%%%%%%%%%%%%%%%%%%
%%%%%%%%%%%%%%%%%%%%%%%%
%INTRODUCTION
%%%%%%%%%%%%%%%%%%%%%%%%
%%%%%%%%%%%%%%%%%%%%%%%%
%%%%%%%%%%%%%%%%%%%%%%%%
\section{Introduction}
Astrophysical simulations of self-gravitating gas often involve regions of gravitational collapse. Resolving those collapses while still following the large scale evolution of the gas therefore requires a huge dynamic range in the density. The local free fall time $t_{\text{ff}}=\sqrt{3\pi/32G\rho}$ is a good estimate for the relevant timescales of the dynamics at a given density. For example, a density contrast of $10^{10}$ observed in giant molecular clouds from the entire cloud down to the first hydrostatic core \citep{Stahler2005} translates into a factor $10^5$ between the smallest and the largest timescale of the problem. Advancing the whole simulation at the smallest time step therefore lets the large scale motions appear completely frozen. Adaptive time stepping that allows for different resolution elements to be updated with different time steps (see \cite{Bate1995} for a description in SPH, \cite{Teyssier2002} for AMR) increases the computationally achievable dynamic range in timescales, but long term evolution of systems hosting sites of gravitational collapse is still not possible in many cases. In addition to the problem of time scales, following the collapsing regions to higher densities requires an ever increasing spatial and mass resolution which increases the necessary number of resolution elements in the simulation. It is therefore inevitable to define a maximum resolution at which one does not follow the ongoing collapse any further. Introducing a maximum resolution raises another problem: as \cite{Truelove1997} have shown, not resolving the Jeans length and Jeans mass in regions of gravitational collapse can lead to artificial fragmentation of the gas. A possible way to avoid this is changing the physical model in a way that will artificially stop the gravitational collapse at a scale that can still be resolved. This is usually achieved by implementing a barotropic equation of state that strongly heats the gas once a certain density is exceeded. \cite{Federrath2010} named this approach `Jeans heating'. A problem of this approach is that objects are kept artificially big and therefore more vulnerable to disruption through shocks and tidal stripping. Another way to deal with limited resolution in simulations of gravitational collapse are sink particles. Instead of artificially stopping the collapse at a chosen scale, sink particles approximate the unresolved small-scale evolution by an immediate collapse onto a point mass. A sink interacts with the remaining gas through gravity and accretion only. Once formed, it is disconnected from the hydrodynamic evolution of the system and can not be destroyed anymore. 

Despite the radical approximations that come with the introduction of a sink particle, they are widely used in simulations of star formation and sink particle schemes are implemented nowadays in many simulation codes. Given the Lagrangian nature of sink particles they have first been introduced in smoothed particle hydrodynamics (SPH) codes  \citep{Gingold1977,Lucy1977}. It was \cite{Bate1995} who presented the first implementation which most subsequent implementations in SPH codes are based upon, like in the codes \textsc{gadget} \citep{Jappsen2005}, \textsc{gasoline}, \citep{Shen2006}, \textsc{dragon}, \citep{Goodwin2004} and \textsc{seven} \citep{Hubber2011}). More recently, \cite{Hubber2013} have introduced a more advanced algorithm that deviates quite strongly from the original one by \cite{Bate1995}. 

\cite{Krumholz2004} were the first to introduce sink particles in the Eulerian, grid-based code \textsc{orion}, built upon the adaptive mesh refinement (AMR) technique \citep[][]{Berger1984,Berger1989}. Their implementation has been the role model for sink particle implementations into various other grid-based codes, such as \textsc{enzo} \citep{Wang2010}, \textsc{ramses} \citep{Dubois2010}, \textsc{pencil} \citep{Padoan2011} and \textsc{orion2} \citep{Lee2014}. Later \cite{Federrath2010} presented their sink implementation into the \textsc{flash} code which deviates considerably from the original \cite{Krumholz2004} method. A more recent implementation has been presented by \cite{Gong2013} for the \textsc{athena} code, quite close to the \cite{Federrath2010} method. While sinks have been used in different context, such as formation and growth of black holes, most of the implementations mentioned above are targeting star formation as the primary application for sink particles.

Simulations of star formation have made tremendous progress throughout the last decade. The increase in computational power and the ongoing evolution of algorithms has allowed simulations of larger volumes and finer resolution. Beyond that, the implementation of radiative transfer, magnetic fields, outflows and chemical evolution models has led to a much better understanding of star formation \citep[e.g.,][]{Offner2009, Wang2010, Bate2012, Krumholz2012}. Some of this additional physics is tightly coupled to the sink particles as they act as a source for feedback processes. This increases the impact of sink particles on the remaining gas. Furthermore, various sink properties such as their mass function, accretion rates, multiplicity fractions and formation rates are used directly for comparison with observations. It is therefore crucial to have reliable sink particle algorithms as well as a good understanding of how the details in the implementation affect the results. 

This is precisely the goal of the present paper: we describe a new, possibly better sink particle implementation together with a suite of test cases that we use for comparing the components of our new algorithm to already existing implementations, mostly in AMR codes. The main novelties in our code are related to the formation of sink particles. We run a clump finder to identify well defined density peaks in the gas which are then treated as possible locations for sink formation. We introduce more exact criteria to check whether the gas inside a small volume around  such a peak is undergoing gravitational collapse and therefore allowed for sink formation. The paper is structured as follows: in Section \ref{sink_creation} we present our algorithm for sink formation and discuss differences and similarities to existing codes. In section \ref{sink_merging} we briefly discuss the issue of sink merging. Section \ref{sink_integration} deals with numerical methods for the integration of the sink particle trajectories. In Section \ref{sink_accretion} we describe different methods for modeling the accretion of gas onto the sink particles. Finally, Section \ref{tests} describes the test and comparison cases that we used to test sink formation, sink merging and the accretion onto sink particles. The Appendix contains a comparison of two integration schemes for the sink particles.

%%%%%%%%%%%%%%%%%%%%%%%%
%%%%%%%%%%%%%%%%%%%%%%%%
%%%%%%%%%%%%%%%%%%%%%%%%
%SINK PARTICLE CREATION
%%%%%%%%%%%%%%%%%%%%%%%%
%%%%%%%%%%%%%%%%%%%%%%%%
%%%%%%%%%%%%%%%%%%%%%%%%
\section{Sink Particle Creation}
\label{sink_creation}
The existing implementations of sink particles into AMR codes can be divided into two classes, namely `cell-based' and `peak-based' techniques. In cell-based methods, sink particles are formed based on purely local quantities. By local, we mean gas properties associated to the corresponding cell only. For example, \cite{Krumholz2004} form sinks in every cell with convergent velocity field whose density exceeds a given threshold. This often results in a connected region where every cell forms a sink particle. These sink particles are then merged using a friends-of-friends (FOF) algorithm \citep{Davis1985}. In contrast, peak-based techniques define small volumes around density peaks above a given density, and apply criteria for sink formation based on quantities integrated over such a volume. This `control volume' around a density peak is usually a sphere with radius chosen equal to the accretion radius \citep{Federrath2010}.

Sink particles inevitably introduce a level of discretisation in our continuous fluid description. Cell and peak-based methods can be seen as different approaches to perform this discretisation. Cell-based approaches form sinks in a continuous, or cell-by-cell way. The discretisation is introduced later by the FOF algorithm, that will break up connected regions into multiple FOF groups. The resulting distribution of the sinks therefore critically depends on the adopted linking length. Peak-based methods introduce discretisation in our fluid by considering only density peaks for sink formation. Note that accretion can affect the results of that procedure by creating a `hole' around the sink and thus creating new artificial sink formation sites close to the boundary of the accretion zone. 

Our new method that we label as `clump-based' is an extension of the peak-based method. Instead of considering every density peak for sink formation, including possibly small fluctuations, we require the peak to have a certain {\it prominence}\footnote{The criterion that we apply is closely linked to the definition of the term {\it prominence} in topography, see Section~\ref{clump_finder} for more details on the clump finder}. Peaks that fail this criterion are considered as `noise' and are merged to neighbouring ones. This provides a more robust segmentation of the volume into a discrete set of subregions, excluding small density fluctuations from the analysis. We consider this as being particularly important if sinks are not allowed to merge during the course of the simulation (see Section \ref{sink_merging} for more details on sink merging). As in the peak-based approach, we define spherical regions around the candidate locations for sink formation. Those regions are then examined for conditions of gravitational collapse.

This raises the question about the size of the region that should be considered. At first sight, taking the accretion zone (i.e. a sphere of radius $R_{\text{acc}}\approx 4\Delta x_{\text{min}}$) as the integration domain for further energetic considerations appears as a natural choice, as it contains the gas from which the sink will form. Considering a larger volume might detect gravitational collapse which can still be well resolved by the simulation and therefore should not trigger sink formation yet. Using a smaller volume leads to a poor definition of quantities such as the internal kinetic energy of the gas inside the sphere. In terms of recent theoretical developments on the origin of the IMF \citep{Hennebelle2008, Hopkins2012}, one can say that the sink particle is introduced when the smallest gravitationally bound scale (`last crossing scale') is of the order of the accretion radius. If we pick the sink formation threshold $\rho_{\text{sink}}$ in agreement with the \cite{Truelove1997} criterion such that the minimum Jeans mass is resolved by 4 cells at the maximum level of refinement, gravity should start to dominate pressure at the scale of the accretion radius which again justifies the use of a sphere of that size to evaluate gravitational collapse. 
As we have just mentioned, the minimum grid spacing sets the maximum density in the simulation (or vice-versa). The remaining free parameter can be set by computational or physical arguments. One can simply choose a certain resolution with respect to the computational resources at hand, knowing that one will miss fragmentation into objects smaller than that scale. Another option is to look for a physical scale (such as the opacity limit in molecular gas at $\sim 10^{-13} \text{g/cm}^3$) to set a minimum scale for fragmentation.

We will now turn to the more detailed description of our new method for sink formation. It consists of the following steps which are described in the following subsections: We check for the creation of new sink particles after every coarse time step.\footnote{\textsc{ramses} allows adaptive time stepping for cells at different levels. This is achieved by updating a fine cell twice while a coarse cell is updated once with a time step which equals the sum of the two fine-level time steps. After every coarse time step, all the levels are synchronized.} First, we run the clump finder to identify peaks and their associated regions. The peak locations identified by the clump finder are taken into account as possible locations $\boldsymbol x_i$ for sink formation. For each of these locations we define a region $\Omega_i$ containing all the cells that lie within the accretion radius from the location considered. The gas inside $\Omega_i$ must be undergoing contraction along \emph{all} directions in order trigger sink formation (collapse check). Furthermore, the gravitational field must be strong enough to overcome all internal support in the gas (virial check). If a peak lies within the accretion radius from a pre-existing sink particle, it is not allowed to form a sink (proximity check).

%%%%%%%%%%%%%%%%%%%%%%%%
%%%%%%%%%%%%%%%%%%%%%%%%
%CLUMPFINDER
%%%%%%%%%%%%%%%%%%%%%%%%
%%%%%%%%%%%%%%%%%%%%%%%%
\subsection{The RAMSES Clump Finder} 
\label{clump_finder}
\begin{figure}
\begin{minipage}[b]{0.49\linewidth}
\includegraphics[scale=0.76]{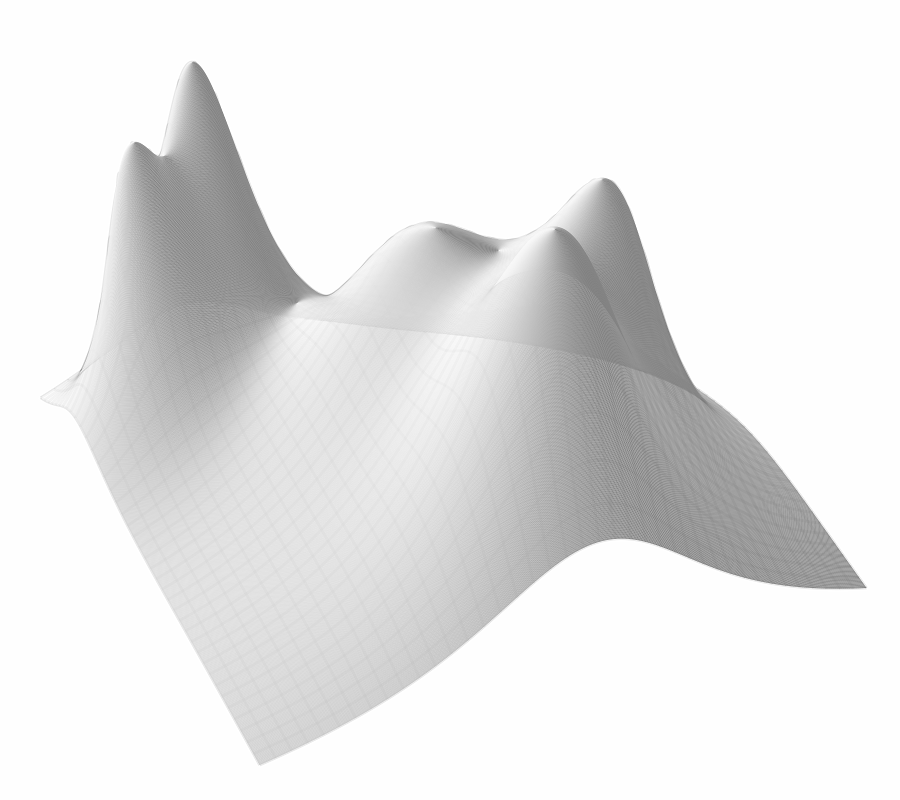}
\caption{}
\label{fig:clfig1}
\end{minipage}
\begin{minipage}[b]{0.49\linewidth}
\includegraphics[scale=0.76]{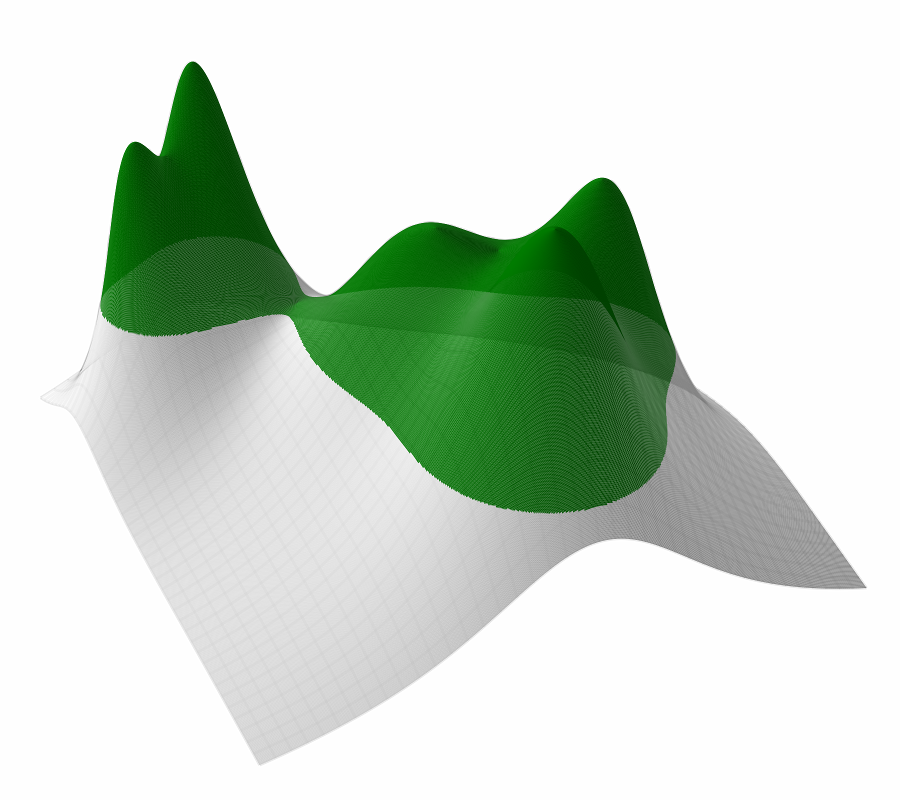}
\caption{}
\label{fig:clfig2}
\end{minipage}
\begin{minipage}[b]{0.49\linewidth}
\includegraphics[scale=0.76]{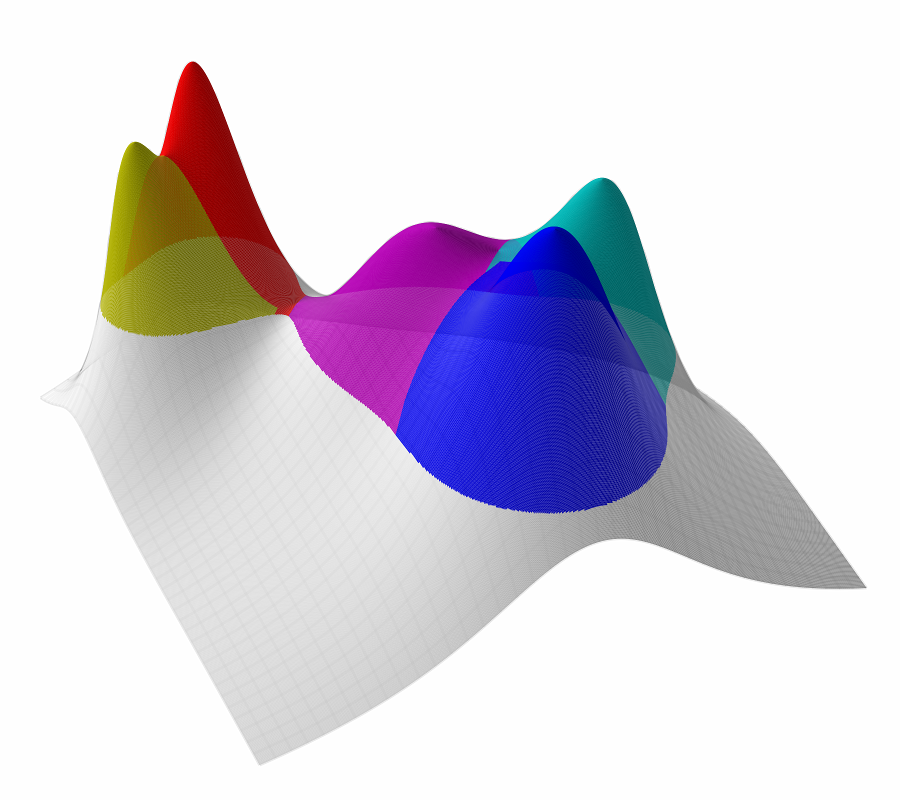}
\caption{}
\label{fig:clfig3}
\end{minipage}
\begin{minipage}[b]{0.49\linewidth}
\includegraphics[scale=0.76]{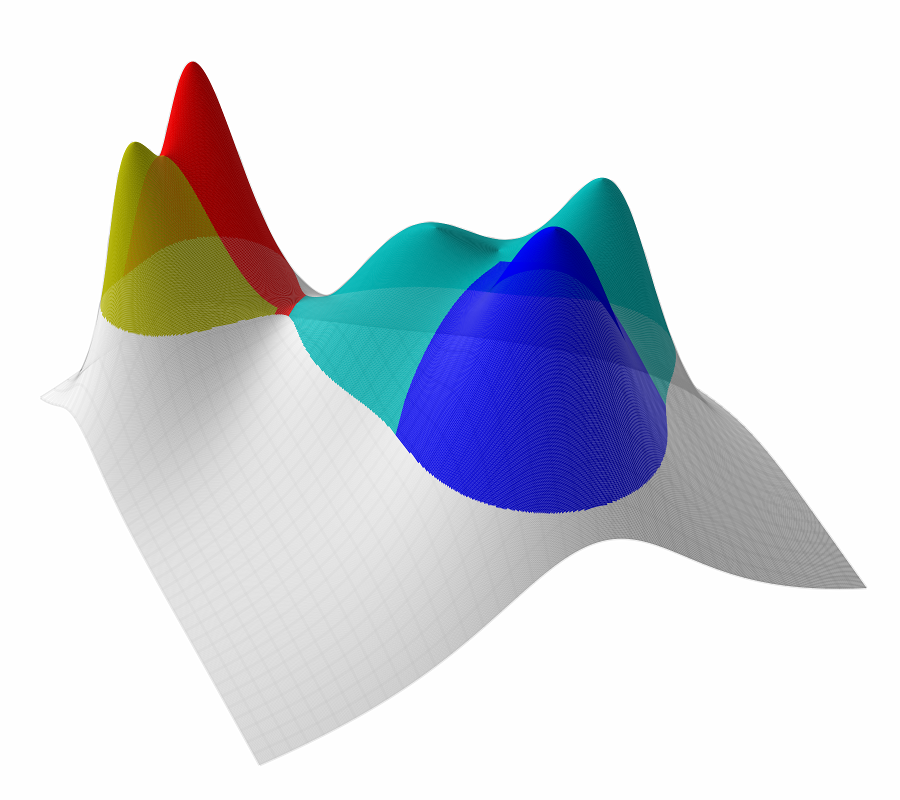}
\caption{}
\label{fig:clfig4}
\end{minipage}
\begin{minipage}[b]{0.49\linewidth}
\includegraphics[scale=0.76]{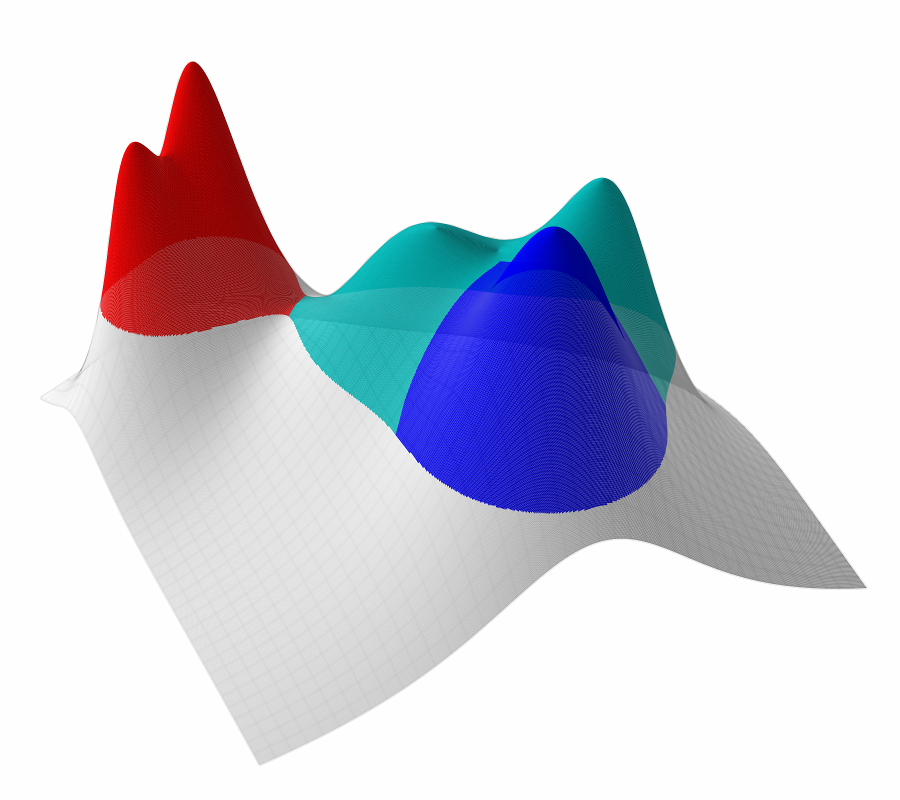}
\caption{}
\label{fig:clfig5}
\end{minipage}
\begin{minipage}[b]{0.49\linewidth}
\includegraphics[scale=0.76]{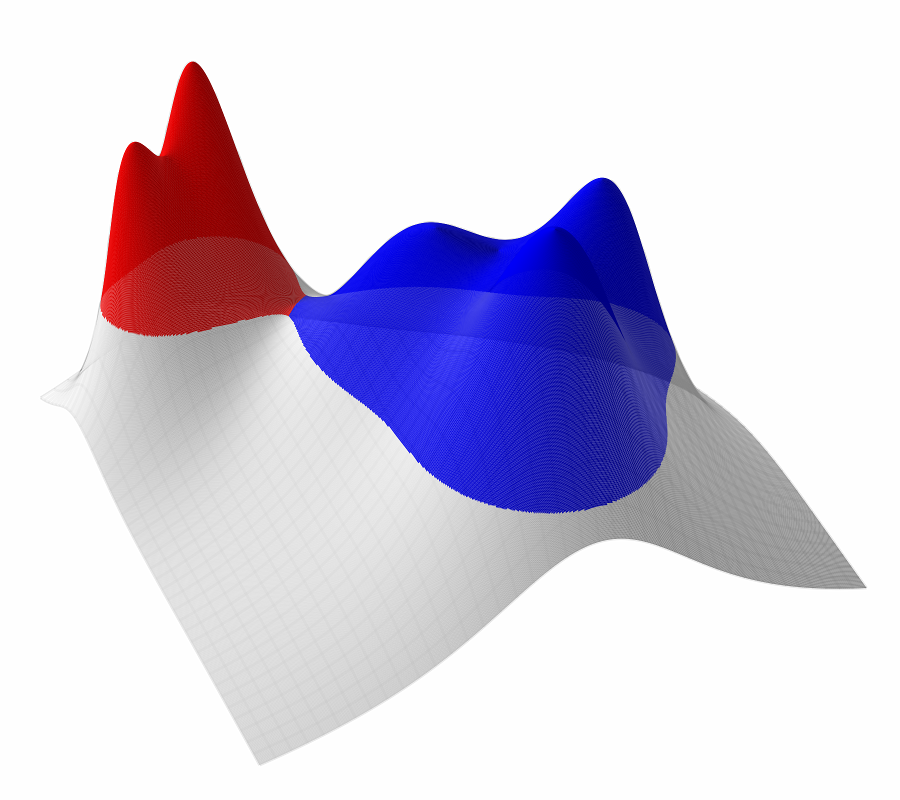}
\caption{}
\label{fig:clfig6}
\end{minipage}
{\bf Figure \ref{fig:clfig1} - \ref{fig:clfig6}.} Working principle of the clump finder represented on a 2d-surface.
\end{figure}

Observers have been identifying bound structure in molecular clouds for a long time\footnote{Finding dark matter haloes in cosmological simulations has been also developed for many decades, and is very similar to finding clumps in turbulent gas. Techniques used in halo finders have influenced our clump finder and can be found in various codes such as \textsc{subfind} \citep{Springel2001} or \textsc{adaptahop} \citep{Aubert2004}.}. \cite{williams1994} describe an algorithm called \textsc{clumpfind} which finds clumps in a PPV (position-position-velocity) cube using a set of isodensity contours. In this method, a gas clump is identified as such if its highest saddle point\footnote{In topography this would be called a key col or key saddle.} is separated from the peak by a contour surface. When operating in log-space with equally spaced contour levels, the contour levels differ by a constant factor in linear space. A clump with a peak-to-saddle ratio above that factor will therefore always be recognized as an individual clump. However, peaks with a lower peak-to-saddle ratio can be separated from their highest saddle point if a contour level happens to be in between the peak and the highest saddle point. Our \textsc{ramses} clump finder defines clumps in a very similar way as the method by \cite{williams1994}. The main difference is that we remove the probabilistic element that comes with the introduction of a finite set of contour levels. Instead of contouring the dataset we identify all peaks and their highest saddle points above a given threshold. We then require the peak-to-saddle ratio to be above a certain value for the peak to survive. Otherwise it is merged to the neighbor it shares the highest saddle point with. We now describe our clump finder in more detail. It works by performing the following steps which are sketched in Figures \ref{fig:clfig1} - \ref{fig:clfig6}: 

\begin{enumerate}

\item In a first step, every cell whose density is higher than a given threshold is marked (Figure \ref{fig:clfig2}). 

\item Every marked cell is then assigned to a density peak by following the path of steepest ascent. We do this by first checking for every marked cell whether it is a local density maximum.\footnote{Note that we consider every cell with a common face, edge or corner as a neighbor of a given cell.} The found maxima are labeled with a global peak-id. All cells above the threshold are sorted in descending density. Next, a loop over all cells is performed where every cell is assigned the peak-id of its densest neighbor. The previous sorting guarantees that the densest neighbor does already have a peak-id assigned. All cells sharing the same peak-id form a so called `peak patch' (Figure \ref{fig:clfig3}). 

\item The saddle point densities connecting between all peak patches are identified. For this purpose we introduce a sparse, symmetric connectivity matrix $M$ of virtual size $n_{\text{peak}}^2$. The value $M(i,j)$ contains the maximum saddle point density connecting peak $i$ with peak $j$. In order to construct this matrix we check for each cell belonging to a certain clump whether it has a neighbor which belongs to a different clump. If this is the case, the average density of the cell and its neighbor is considered the density at the common surface and written into the $M(i,j)$ if it is bigger than the existing value. The highest saddle point lying on the boundary of a certain peak patch is the relevant one for our analysis. This corresponds to the maximum of a certain line in the connectivity matrix. By looking at the ratio of the peak density to the maximum saddle density of a peak we decide whether this is a significant one or not. We usually require this peak-to-saddle ratio to be bigger than 2.\footnote{
The exact choice of this value is not critical for the formation of sink particles. The checks which are applied later (see Section \ref{sink_creation}) usually ensure a higher peak-to-saddle ratio than what we require here.} 
\item The peak patches are sorted by ascending peak density. Insignificant peak patches are merged to the one they are connected to through the highest saddle point. The sorting is important since it makes sure that no peak patch is merged with one that has already been merged into another one before. Isolated peak patches which are insignificant are rejected (Figures \ref{fig:clfig4},\ref{fig:clfig5}). After every single merger, we update the connectivity matrix and the peak-to-saddle ratio of the peak patch that has grown due to the merger.
\item After the previous step all insignificant peak patches have been rejected or merged to form significant ones which we now label as clumps (Figure \ref{fig:clfig6}). The list of mergers is used to link every peak patch initially present (Figure \ref{fig:clfig3}) to the final clump in the merging history and all cells above the density threshold are reassigned their new peak-id.
\end{enumerate}

Since we want to use our clump finder to find possible locations for sink particle creation, it needs to run on the fly. It is therefore implemented in a parallel fashion. The steps (i-v) need to be adapted in order to the be implemented in a MPI code where every MPI domain only contains a fraction of the whole computational domain. In \textsc{ramses} \citep{Teyssier2002} the cells that belong to a MPI processes domain (`active' cells) are wrapped in a thin layer of cells that do belong to neighboring MPI domains (`virtual boundaries'). In step (i) only active cells are flagged. In step (ii) the flagged cells are sorted inside each MPI domain and the loop over all cells is performed by each MPI process individually. After this loop, the peak-id of active cells close to a domain boundary are copied into the virtual boundary regions of the neighboring MPI domains and the loop is repeated until every cell is either a local maximum or has the same peak-id as its densest neighbor. In step (iii) we keep the connectivity matrix $M(i,j)$ local to each domain, while the other quantities of the peaks (peak density, peak position) are global in the sense that all MPI processes have the information about all peaks. When clump $i$ needs to be merged, every MPI process searches for its own maximum in the $i$-th line of $M(i,j)$. The values of all the maxima are compared between the MPI processes to find the index of the global maximum. The mergers in step (iv) and the final link from initial peak-id to final peak-id in step (v) are performed globally by all MPI processes and the actual reassignment of cells with their final peak-id is done by each MPI process for its active cells.

%%%%%%%%%%%%%%%%%%%%%%%%
%%%%%%%%%%%%%%%%%%%%%%%%
%VIRIAL CHECK
%%%%%%%%%%%%%%%%%%%%%%%%
%%%%%%%%%%%%%%%%%%%%%%%%
\subsection{Virial Check}\label{virial_check}
The gas surrounding the density peaks found by the clump finder is investigated for gravitational collapse. We perform a virial theorem type analysis to balance the gas configurations self gravity against the gas internal support. As it is done in textbooks when deriving the virial theorem \citep[e.g.,][]{Stahler2005} we start by defining the scalar moment of inertia 
\begin{equation}\label{scalar_moment}
	\mathcal I=\int_{\Omega_i} \rho |\boldsymbol r|^2 \mathrm d V
\end{equation}
as a measure of the spatial extent of the gas configuration contained in $\Omega_i$. The corresponding acceleration is found computing the second derivative in time of $\mathcal I$. Since the volume $\Omega_i$ is moving with the flow, we apply Reynolds transport theorem twice to obtain
\begin{equation}\label{scalar_moment2}
	\frac 1 2 \frac {\mathrm d^2}{\mathrm d t^2} \mathcal{I}= \int_{\Omega_i}  \rho |\boldsymbol v|^2 \mathrm d V+\int_{\Omega_i}\rho \bigg( \boldsymbol r  \cdot \frac {\mathrm D \boldsymbol v}{\mathrm D t}\bigg)  \mathrm d V,
\end{equation}
where the $ \mathrm D / \mathrm D t $ operator stands for the Lagrangian derivative. We now write the Euler equation in Lagrangian form, using gravitational and radiative acceleration as external forces and the general form of the stress tensor $\overline {\overline {\sigma}}$ for internal forces,
\begin{equation}\label{euler}
	\rho \frac {\mathrm D \boldsymbol v}{\mathrm D t}=\rho \boldsymbol g + \frac{\kappa \rho}{c} \boldsymbol F_{\text{rad}}+ \nabla \cdot \overline {\overline {\sigma}}.
\end{equation}
In the previous equation $\boldsymbol g$ stands for the gravitational acceleration and $\boldsymbol F_{\text{rad}}$, $\kappa$, $c$ are the radiation flux the opacity and the speed of light. Injecting the Euler equation in Equation \ref{scalar_moment2} gives

\begin{multline}\label{full_virial}
	\frac 1 2 \frac {\mathrm d^2}{\mathrm d t^2} \mathcal I = \int_{\Omega_i} \rho |\boldsymbol v|^2 \mathrm d V+
	\int_{\Omega_i} \rho \, \boldsymbol g \cdot \boldsymbol r \, \mathrm d V\\
	+\int_{\Omega_i} \frac{\kappa \rho}{c} \mathrm{r} \cdot \boldsymbol F_{\text{rad}}  \mathrm d V+
	\int_{\Omega_i}\mathrm{r} \cdot ( \nabla \cdot \overline {\overline {\sigma}})\, \mathrm d V.
\end{multline}
We use the vector identity

\begin{equation}\label{vector_id}
	\nabla \cdot ( \overline {\overline {\sigma}} \boldsymbol r)= \boldsymbol r \cdot (\nabla \cdot \overline {\overline {\sigma}} ) + \mathrm{Tr}( \overline {\overline {\sigma}} )
\end{equation}
to obtain the virial theorem in its generalized form,

\begin{multline}\label{final_virial}
	\frac 1 2 \frac {\mathrm d^2}{\mathrm d t^2} \mathcal I = \int_{\Omega_i} \rho |\boldsymbol v|^2 \mathrm d V+
	\int_{\Omega_i} \rho \, \boldsymbol g \cdot \boldsymbol r \, \mathrm d V\\
	+\int_{\Omega_i} \frac{\kappa \rho}{c} \mathrm{r} \cdot \boldsymbol F_{\text{rad}}  \mathrm d V-
	\int_{\Omega_i} \mathrm{Tr}( \overline {\overline {\sigma}})  \, \mathrm d V +
	\int_{\partial  \Omega_i}\boldsymbol r \cdot (\overline {\overline {\sigma}} \boldsymbol  n )\, \mathrm d A.
\end{multline}
We have used the divergence theorem to transform the volume integral over the left-hand term in Equation \ref{vector_id} into a surface integral, $\partial  \Omega_i$ therefore denotes the boundary of $ \Omega_i$ and $\boldsymbol n$ is the outward pointing unit normal to the boundary. The stress tensor can be written in general for a viscous magnetized fluid as

\begin{equation}\label{stress}
	\overline {\overline {\sigma}} = - P \mathbb{I}+\overline{\overline {\tau}}+\overline {\overline {\mathcal M}} 
\end{equation}
where $\overline {\overline {\tau}}$ is the viscous stress tensor and the magnetic stress is given in the ideal MHD limit by the Maxwell tensor

\begin{equation}\label{maxwell}
	\mathcal M_{ij}=B_i B_j - \frac {B^2}{2} \delta_{ij}.
\end{equation}
In the case of isotropic stresses and without radiation, this simplifies into

\begin{multline}\label{Idotdot}
	\frac 1 2 \frac {\mathrm d^2}{\mathrm d t^2} \mathcal I = 
	\underbrace{\int_{\Omega_i} \rho |\boldsymbol v|^2 \mathrm d V}_{\text{kinetic energy term}}
	+\underbrace{\int_{\Omega_i} \rho \, \boldsymbol g \cdot \boldsymbol r \, \mathrm d V}_{\text{tidal energy term}}\\
	+\underbrace{\,3\int_{\Omega_i} P \mathrm d V }_{\text{volume pressure term}}
	-\underbrace{\int_{\partial  {\Omega_i}}P \, \boldsymbol n \cdot \boldsymbol r \, \mathrm d A}_{\text{surface pressure term}}, 
\end{multline}
While the first term on the right-hand side is indeed twice the kinetic energy, one must make further assumptions if one wishes to simplify this into a more common form of the virial theorem.
The second term is usually identified as the total gravitational energy
\begin{equation}\label{E_pot}
	E_{\text{pot}}=\frac 1 2 \int_{\Omega_i} \rho\, \phi_g \mathrm d V,
\end{equation}
This is valid only if the potential is caused entirely by the gas inside $\Omega_i$. 
A physically more correct interpretation of this term is obtained using a first order Taylor expansion of the gravity acceleration with respect to the center of mass as
\begin{equation}
\boldsymbol g \simeq \boldsymbol g_{cm} + \overline {\overline {\cal T}} \left( \boldsymbol r - \boldsymbol r_{cm}\right)
\end{equation}
where $\overline {\overline {\cal T}}$ is the {\it tidal tensor}, so that the tidal energy term can be written to leading order as
\begin{equation}
\int_{\Omega_i} \rho \, \boldsymbol g \cdot \boldsymbol r \, \mathrm d V \simeq 
\boldsymbol g_{cm} \cdot \boldsymbol r_{cm} + 
\int_{\Omega_i} \rho \, \boldsymbol r_{rel} \cdot  \overline {\overline {\cal T}} \boldsymbol r_{rel} \, \mathrm d V
\end{equation}
\noindent
which demonstrates that this term is related to the tidal tensor, not to the potential energy. 
The third term is equal to the thermal energy only for certain equations of state. 
Furthermore, the pressure surface term is often neglected. These various approximations might be justified when considering a gas configuration which is (nearly) in isolation. 
However, the gas from which sinks form is typically far from being isolated and we therefore do not simplify Equation \ref{Idotdot} any further.

Just as the inertia tensor that arises when studying the dynamics of rigid bodies, the scalar moment of inertia defined in Equation \ref{scalar_moment} depends on the choice of the coordinate system. Starting from the scalar moment of inertia in the center of mass frame, $\mathcal I _{\text{cm}}$, we find $\mathcal I$, the scalar moment of inertia of the same object with center of mass located at position $\boldsymbol x_{\text{cm}}$, using the equivalent of the parallel-axis theorem

\begin{equation}\label{steiner}
	\mathcal I=\mathcal I_{\text{cm}}+M |\boldsymbol r_{\text{cm}} |^2.
\end{equation} 
While the above derivations of the generalized virial theorem hold for any inertial frame of reference, the interpretation of $\mathcal{I}$ as a measure of the size of the gas configuration only makes sense as long at the coordinate origin is located in the center of mass. We therefore choose the frame which is comoving with the center of mass of the gas contained in $\Omega_i$. This non-inertial frame gives rise to a fictitious acceleration that enters Equation \ref{euler}. As long as the frame is non-rotating, this additional acceleration term is independent of the position in space and its contribution to the second term on the right-hand side of Equation \ref{final_virial} vanishes in the center of mass frame\footnote{
For a spatially constant fictitious acceleration $\boldsymbol g_{\text{fict}}$ we have 

\begin{equation}\label{cond}
	\int_{\Omega_i} \rho \, \boldsymbol g_{\text{fict}} \cdot \boldsymbol r \, \mathrm d V={\boldsymbol g}_{\text{fict}} \cdot \int_{\Omega_i} \rho \, \boldsymbol r \, \mathrm d V= {\boldsymbol g}_{\text{fict}} \cdot \boldsymbol r_{\text{cm}} 
\end{equation}
which vanishes since $\boldsymbol r_{\text{cm}}=0$ in the center of mass frame.}.
We thus rewrite Equation \ref{final_virial} in the center of mass frame

\begin{multline}\label{final_virial_rel}
	\frac 1 2 \frac {\mathrm d^2}{\mathrm d t^2} \mathcal I_{\text{cm}} = \int_{\Omega_i} \rho |\boldsymbol v_{\text{rel}}|^2 \mathrm d V+
	\int_{\Omega_i} \rho \, \hat {\boldsymbol g}_{\text{rel}} \cdot \boldsymbol r_{\text{rel}} \, \mathrm d V\\
	-\int_{\Omega_i} \mathrm{Tr}( \overline {\overline {\sigma}})  \, \mathrm d V +
	\int_{\partial  \Omega_i}\boldsymbol r_{\text{rel}} \cdot (\overline {\overline {\sigma}} \boldsymbol  n )\, \mathrm d A.
\end{multline}
where the index `rel' refers to the position, velocity and acceleration relative to their centre of mass values. For simplicity we have absorbed the radiation force as an effective gravitational acceleration

\begin{equation}\label{g_eff}
	\hat {\boldsymbol g}=\boldsymbol g+\frac \kappa c \boldsymbol {F}_{\text{rad}}.
\end{equation} 
It is this last version of the virial theorem that we use as check for sink formation. For an inviscid gas in the absence of radiation and magnetic fields as it is the case in the tests described in Section \ref{tests}, $\hat {\boldsymbol g}_{\text{rel}}$ is therefore simply the relative gravitational acceleration $ \boldsymbol g_{\text{rel}}$ and the stress is given by $\overline {\overline {\sigma}}= - P \mathbb{I}$. Equation \ref{final_virial_rel} simplifies to 
\begin{multline}\label{final_virial_rel2}
	\frac 1 2 \frac {\mathrm d^2}{\mathrm d t^2} \mathcal I_{\text{cm}} = \int_{\Omega_i} \rho |\boldsymbol v_{\text{rel}}|^2 \mathrm d V+
	\int_{\Omega_i} \rho \, {\boldsymbol g}_{\text{rel}} \cdot \boldsymbol r_{\text{rel}} \, \mathrm d V\\
	+3\int_{\Omega_i} P \mathrm d V 
	-\int_{\partial  {\Omega_i}}P \, \boldsymbol n \cdot \boldsymbol r_{\text{rel}} \, \mathrm d A,
\end{multline} 
which is the same as Equation \ref{Idotdot}, but this time in the comoving, non-inertial center of mass frame. Note that the last term in the above equation simplifies to $4\pi R^3 P_{\text{surface}}$ for a spherical region of radius R, which cancels with the volume pressure term in the case of constant pressure. The gas in $\Omega_i$ is only further considered for sink formation, if $ \ddot{\mathcal I}_{\text{cm}}<0$. This condition ensures that the gravitational field at a possible location for sink formation is compressive and strong enough to overcome all internal support present in the gas. In contrast to estimations of the gravitational potential energy that do neglect the curvature of the background potential, our version fully takes into account any tidal forces that could prevent the collapse of the gas. All the required quantities are readily available in the computational code, which makes this condition well suited for implementation in simulations.

%%%%%%%%%%%%%%%%%%%%%%%%
%%%%%%%%%%%%%%%%%%%%%%%%
%COLLAPSE CHECK
%%%%%%%%%%%%%%%%%%%%%%%%
%%%%%%%%%%%%%%%%%%%%%%%%
\subsection{Collapse Check}\label{collapse_check}
The gas which is about to form a sink particle must not only be accelerated towards the center of the volume under consideration, it must as well be contracting at the moment of formation. \cite{Krumholz2004} require $\nabla \cdot \boldsymbol v <0$ for a cell which is about to form a sink. \cite{Federrath2010} apply a similar check by requiring that the gas inside the `control volume' is contracting along all principal axes. We adapt this criterion to our analysis presented in Section \ref{virial_check} and compute all eigenvalues $\lambda_1,\lambda_2,\lambda_3$ and the corresponding normalized eigenvetors $\boldsymbol e_1,\boldsymbol e_2,\boldsymbol e_3$ of the symmetric tensor \footnote{Note that 

\begin{equation}\label{cont_cond2}
	({\overline{\overline{I}}_\text{cm}}\boldsymbol u)\cdot \boldsymbol u=\int_{\Omega_i} \rho \, (\boldsymbol u \cdot \boldsymbol r_{\text{rel}})^2 \, \mathrm d V.
\end{equation}
is a measure of the extension of an object along a certain direction specified by the unit vector $\boldsymbol u$.}

\begin{equation}\label{cont_cond3}
	{\overline{\overline{I}}_\text{cm}}=\int_{\Omega_i} \rho \, \boldsymbol r_{\text{rel}} \otimes \boldsymbol r_{\text{rel}} \, \mathrm d V.
\end{equation}
By computing the time derivative 

\begin{equation}\label{cont_cond4}
	\frac {\mathrm d \overline{\overline{I}}_\text{cm}}{\mathrm d t} = \int_{\Omega_i} \rho ( \boldsymbol r_{\text{rel}} \otimes \boldsymbol v_{\text{rel}} + \boldsymbol v_{\text{rel}} \otimes \boldsymbol r_{\text{rel}}) \mathrm d V
\end{equation}
we can assign a collapse timescale to each direction given by the eigenvectors of  ${\overline{\overline{I}}_\text{cm}}$

\begin{equation}\label{cont_timescale}
	t_i=\frac{\lambda_i}{(\frac {\mathrm d \overline{\overline{I}}_\text{cm}}{\mathrm d t} \boldsymbol e_i)\cdot \boldsymbol e_i},
\end{equation}
where a small negative timescale indicates fast collapse along a certain axis. Only one negative timescale is a sign for sheet-like and two negative timescales indicate filamentary collapse. Although collapsing, these collapsed regions are poorly approximated by a point mass. We therefore require all three timescales to be negative to ensure collapse onto a point-like object before we introduce a sink particle. This condition can be further strengthened by enforcing collapse along all axis within a certain time (see \ref{sink_merging_timescale}).

%%%%%%%%%%%%%%%%%%%%%%%%ø
%%%%%%%%%%%%%%%%%%%%%%%%
%PROXIMITY CHECK
%%%%%%%%%%%%%%%%%%%%%%%%
%%%%%%%%%%%%%%%%%%%%%%%%
\subsection{Proximity Check}\label{proximity_check}
Gas which is falling onto an existing sink particle is not allowed to form another sink, even if there is a density peak which fulfills all criteria for sink formation. We therefore check whether the possible location is closer than one accretion radius from an existing sink. If it is, we do not allow formation of a new sink. \cite{Federrath2010} applied this test that can be seen as the possibility for sinks to merge to existing ones at their time of birth (see Section \ref{sink_merging}).

%%%%%%%%%%%%%%%%%%%%%%%%
%%%%%%%%%%%%%%%%%%%%%%%%
%FURTHER CHECKS
%%%%%%%%%%%%%%%%%%%%%%%%
%%%%%%%%%%%%%%%%%%%%%%%%
\subsection{Alternative Checks}
We briefly present and discuss alternative checks which we implemented for testing and comparison reasons, but are not used in our final version of the code. All these tests have been described by \cite{Federrath2010} to whom we refer for more details.

\subsubsection{Bound state check}
The total energy in the control volume must be negative to form a sink,
\begin{equation}\label{grav_bound}
	E_{\text{pot}}+E_{\text{kin}}+E_{\text{therm}} + E_{\text{mag}}<0.
\end{equation}
It seems obvious that a sink particle should only be formed out of gas which is gravitationally bound. One can thus call this a necessary condition for gravitational collapse. However, the condition is not sufficient. A gas configuration in virial equilibrium passes this test although it is not collapsing. Furthermore it is not straightforward to define the gravitational binding energy $E_{\text{pot}}$ of a gas configuration which is embedded in a cloud of turbulent gas. When we use this check in our comparison tests, we compute the maximum potential inside $\Omega_i$ and use this as a reference potential.

\subsubsection{Jeans instability check}
The mass inside the control volume must exceed the local Jeans mass. This is made sure by requiring
\begin{equation}\label{jeans_instab}
	E_{\text{pot}}+2E_{\text{therm}}<0.
\end{equation}
As the bound state check, this condition represents a necessary but not a sufficient condition for gravitational collapse as it neglects the internal kinetic energy of the gas and it is not clear how to define $E_{\text{pot}}$.

\subsubsection{Potential minimum check}
\cite{Federrath2010} introduced this check which has been adopted by other groups in AMR \citep{Gong2013} or SPH \citep{Wadsley2011, Hubber2013} codes to reduce the formation of spurious sinks. This check allows a sink to be formed only in a cell which hosts a local minimum in the gravitational potential. Although the authors mentioned above find this test important to reduce the production of sinks from transient density fluctuations, it is lacking of a physical justification. A local minimum in the gravitational potential is not a prerequisite for local gravitational collapse. This can be seen in a thought experiment where a constant force field is applied to the region of interest. The addition of a constant force term corresponds to adding a linear term in the gravitational potential. This changes the position and/or existence of local extrema in the potential without changing the local dynamics. This demonstrates why the tidal tensor, which is not affected by the addition of a linear term, is the right quantity for the evaluation of local gravitational collapse (see Section \ref{virial_check}). It is therefore not clear whether the gravitational potential due to pre-existing sinks should be added to the gas potential before applying this check or not\footnote{
	This question only arises when the direct force summation approach is used. When applying the PM method the sink mass is contained in the source term of the Poisson equation and therefore in the resulting potential (see Section \ref{sink_integration}).
}. 
Including the sink potential introduces strong gradients which could wrongfully prevent a sink from being formed by removing or dislocating the potential minimum. On the other hand, the curvature of the potential induced by the sink particles contains the tidal forces that the sink particles exert onto the surrounding gas and should therefore enter the analysis. In our implementation of the potential minimum check we decided to consider only the gravitational potential caused by the gas.

%%%%%%%%%%%%%%%%%%%%%%%%
%%%%%%%%%%%%%%%%%%%%%%%%
%MERGING SINKS
%%%%%%%%%%%%%%%%%%%%%%%%
%%%%%%%%%%%%%%%%%%%%%%%%
\section{Merging Sinks}\label{sink_merging}
Sink particles are typically introduced to represent gravitationally collapsed objects whose physical size is orders of magnitude below the grid scale. To decide whether two of those objects are undergoing a merger is therefore beyond the scope of the simulation itself, even in cases where the two sink particles are occupying the exact same cell for a long time. We have to consider physics on a sub-grid scale to decide whether two objects which are close to each other relative to the grid scale will actually get close to each other on the scale of their physical extent. Approaches to sink merging in existing implementations therefore cover a broad spectrum. \cite{Krumholz2004} merge sinks using a FOF algorithm where the linking length is given by the accretion radius of the sink. Formation and subsequent merging of sinks can be seen as one mode of accretion. This merging strategy is clearly targeting young sink particles and the authors mention the possibility to turn off merging at a later stage during the simulation. \cite{Wang2010} and \cite{Krumholz2012} have presented calculations where they use a mass threshold which - once a sink particle has passed it - prevents the sink from being destroyed through merging. \cite{Gong2013} follow a merger friendly strategy as well and merge sink particles as soon as their accretion zones are overlapping. \cite{Federrath2010} have implemented sink merging as an option that can be activated by the user. If switched on, two sink particles will merge whenever their separation is less than one accretion radius, they are converging and they are gravitationally bound to each other. 

\subsection{Merging on a Timescale}\label{sink_merging_timescale}
As \cite{Federrath2010} we share the view that sink merging should be optional in a simulation code since it must be decided based on the very details of the setup and the sub grid physics whether sinks should merge or not. However, in order to bridge the gap between the two extreme cases we present a strategy where we merge sinks based on a collapse timescale. The underlying assumption is that the gas which has just triggered sink formation takes a certain time to collapse to sub grid scale. During this time, the sink represents a `not yet collapsed' object whose size is still comparable to the grid spacing. We therefore merge such a young sink to an `old' one if they are less than one accretion radius apart, or we merge two young sinks if their distance is less than two accretion radii.  When we apply this method, we slightly modify the checks for sink creation to be more consistent with the idea of a collapse timescale. In Section \ref{collapse_check} we introduced three timescales of contraction (see Equation \ref{cont_timescale}). For sink creation we therefore require the contraction time scale along each direction to be shorter than the chosen time scale of collapse. While this time scale must be adapted to the physical setup considered, the concept was clearly motivated by the lifetime of the first Larson core in simulations of fragmenting turbulent molecular clouds \citep{Larson1969}.

%%%%%%%%%%%%%%%%%%%%%%%%
%%%%%%%%%%%%%%%%%%%%%%%%
%INTEGRATION
%%%%%%%%%%%%%%%%%%%%%%%%
%%%%%%%%%%%%%%%%%%%%%%%%
\section{Sink Particle Trajectories}\label{sink_integration}
The integration of sink particle motion in different AMR codes mainly differs in the way the sink-sink and sink-gas gravitational forces are computed. A natural approach for a particle mesh code (PM, \cite{Hockney1981}) such as \textsc{ramses} is to use the PM method for the sink particles in a similar way as it is used for dark matter particles. Another option is to compute the sink-sink and sink-gas interactions `brute force' by summing up the pairwise forces directly. Direct summation consists of a loop of size $n_{\text{cells}} \cdot n_{\text{sinks}}$ and one of size $n_{\text{sinks}}^2$. Simulations involving a large number of sink particles and cells can be slowed down so much that switching to the PM method might be desirable. However, the PM method is not designed for collisional dynamics. We expect it to be inaccurate for situations where the local gravitational field is completely dominated by a sink particle. \cite{Federrath2010} use direct force summation for the sink-sink acceleration and the gas acceleration due to the sink, for the sink acceleration due to the gas they perform `cloud-in-cell' (CIC, \cite{Hockney1981}) interpolation of the gravitational field from the grid values onto the location of the sink. \cite{Krumholz2004} do direct force summation as well, while \cite{Gong2013} use the PM method together with the `triangular-shaped-cloud' (TSC, \cite{Hockney1981}) interpolation scheme. Another distinguishing feature of certain sink particle implementations \citep{Krumholz2004, Federrath2010} is the possibility for the sink particles to `sub-cycle' the gas, meaning that multiple sink particle updates are performed within one time step of the computationally much more expensive hydro solver. This technique therefore allows a very small softening length for sink-sink interactions (or no softening at all) which pushes the resolution of the sink-sink forces beyond the grid spacing.

\subsection{PM Method}
Our implementation of the PM method for sink particles makes use of the PM method for dark matter particles already present in \textsc{ramses} \citep{Teyssier2002}. Each sink particles mass is distributed equally onto a spherical `swarm' of equally spaced \textsc{ramses} particles. The spacing of these particles is half the grid spacing, the radius of the sphere is a free parameter and sets the gravitational softening length. The mass of each particle is deposited onto the grid using the CIC scheme with cloud size being equal to the local grid spacing. This can be seen as a `fuzzy' top hat softening. The Poisson equation is solved using one of the solvers implemented in \textsc{ramses} (multigrid: \cite{Guillet2011}, conjugate gradient: \cite{Teyssier2002}) and the gravitational field is computed using the 5 point finite difference approximation. The gravitational acceleration of each swarm particle is obtained by CIC interpolation from the cell center values. Finally, averaging over all particles belonging to one sink yields the acceleration of the sink.

\subsection{Direct Force Summation}
When doing direct force summation, only the gas density is considered as source term for the Poisson equation. Accelerations due to sink-sink and sink-gas interactions are computed by looping over all pairwise combinations and computing their mutual attraction.\footnote{All the gas in one cell is assigned to the cell center location for this step.} We apply a Plummer softening \citep{Aarseth1963}.
\begin{equation}\label{plummer}
	\boldsymbol F (\boldsymbol r)=-\boldsymbol r\frac{GM}{(|\boldsymbol r |^2+R_{\text{soft}}^2)^{3/2}}
\end{equation}
to both, the sink-sink and the sink-gas forces where the softening length is a free parameter. As \cite{Krumholz2004} point out, the gravitational force should not be reduced too much due to the softening at the boundary of the sink accretion zone. We therefore set the softening radius to half the accretion radius as a default. This implies that the resolution of the sink-sink forces is of the order of the grid spacing. We are therefore for instance not able to follow two sinks orbiting each other inside one cell. 

\subsection{The Integrator}
In \textsc{ramses} particles are integrated using a second order midpoint scheme which - for constant time steps - is equivalent to the classical leapfrog method \citep{Teyssier2002}. We apply the same method to the sink particles. Since we use identical softening for sink-sink as for sink-gas forces, the maximum accelerations of gas and sink particles are comparable. We therefore update the sink particles using the same time step as for the gas at the finest level of refinement.\footnote{\textsc{ramses} allows a finer level in the AMR hierarchy to `sub-cycle' a coarser level by updating the finer level twice while the coarse level is updated only once.} In \textsc{ramses} calculations the minimum free fall time occurring has to be resolved, 
\begin{equation}\label{freefall}
	\Delta t < C \sqrt{\frac{3\pi}{32 G \rho_{\text{max}}}},
\end{equation}
where $0<C<1$ is a constant \citep{Teyssier2002}. When using the PM method, the maximum density $\rho_{\text{max}}$ is identified after the particle mass deposition through the CIC scheme. In case of direct force summation, as soon as the maximum sink density obtained from the Plummer density distribution $\rho_{\text{Plummer}}={3M_{\text{sink}}}/{4 \pi r_{\text{soft}}^3}$ exceeds the maximum gas density, $\rho_{\text{Plummer}}$ is used for computing the time step through Equation \ref{freefall}. Furthermore, sink particles like any other particle in \textsc{ramses} are allowed to travel only a fraction of the local mesh spacing within one time step. As a last sink related restriction on the time step, we set the condition that only a fraction of the available gas can be accreted within one time step (see Section \ref{flux accretion}).

%%%%%%%%%%%%%%%%%%%%%%%%
%%%%%%%%%%%%%%%%%%%%%%%%
%ACCRETION ONTO SINKS
%%%%%%%%%%%%%%%%%%%%%%%%
%%%%%%%%%%%%%%%%%%%%%%%%
\section{Accretion onto Sinks}\label{sink_accretion}
After its formation, a sink particle accretes gas from nearby cells. Different methods to perform accretion have been described and justified using various tests. However, direct comparisons of results obtained by different accretion schemes have not been performed. We implemented and compared three different modes of accretion. Fixed threshold accretion (TA), Bondi-Hoyle accretion (BH) and what we call flux accretion (FA) where the accretion rate is computed based on the mass flux rate into the sink accretion zone. In the following subsections we briefly describe the different schemes. In all schemes, velocity and position of the accreted gas relative to the sink are used to update position and velocity of the sink as well as to keep track of the angular momentum that has been removed from the gas by the sink particle.

\begin{equation}\label{acc_mass_cons}
	M_{s}^{\text{new}}=M_{s}^{\text{old}}+\sum_{i \in cells} \Delta m_i
\end{equation}
\begin{equation}\label{acc_loc_cons}
	\boldsymbol{R}_s^{\text{new}}=(\boldsymbol{R}_s^{\text{old}}M_s^{\text{old}}+\sum_{i \in cells} \boldsymbol{r_i} \Delta m_i)/M_{s}^{\text{new}}
\end{equation}
\begin{equation}\label{acc_mom_cons}
	\boldsymbol{V}_s^{\text{new}}=(\boldsymbol{V}_s^{\text{old}}M_s^{\text{old}}+\sum_{i \in cells} \boldsymbol{v_i} \Delta m_i)/M_{s}^{\text{new}}
\end{equation}
\begin{multline}\label{acc_angmom_cons}
	\boldsymbol{L}_s^{\text{new}}=\boldsymbol{L}_s^{\text{old}}+(\boldsymbol{R}_s^{\text{new}}-\boldsymbol{R}_s^{\text{old}})\times(\boldsymbol{V}_s^{\text{new}}-\boldsymbol{V}_s^{\text{old}})M_s^{\text{old}} +\\
	 \sum_{i \in cells}(\boldsymbol{R}_s^{\text{new}}-\boldsymbol{r}_i)\times(\boldsymbol{V}_s^{\text{new}}-\boldsymbol{v}_i) \Delta m_i
\end{multline}

%%%%%%%%%%%%%%%%%%%%%%%%
%FIXED THRESHOLD
%%%%%%%%%%%%%%%%%%%%%%%%
\subsection{Threshold Accretion (TA)}
\cite {Federrath2010} use this method where gas is accreted from cells which are closer than $R_{\text{acc}}$ to an existing sink and whose density exceeds the threshold $\rho_{\text{sink}}$. Additionally, the gas in a cell is required to be bound to the sink and the radial component of the gas velocity relative to the sink needs to be negative. If these conditions are met, the accreted gas mass from a cell is
\begin{equation}\label{acc_mass_thresh}
	\Delta m_i=\max(0.5(\rho-\rho_{\text{sink}})(\Delta x)^3,0),
\end{equation}
where $\Delta x$ is the size of the cell. In sorting the sink particles by mass we ensure that the the most massive sink gets most of the mass in case of multiple sinks accreting from the same cell. \cite{Federrath2010} improve this by checking which sink the gas is bound to the strongest.

%%%%%%%%%%%%%%%%%%%%%%%%
%BONDI ACCRETION
%%%%%%%%%%%%%%%%%%%%%%%%
\subsection{Bondi-Hoyle Accretion (BH)}\label{Bondi-Hoyle accretion}
\cite{Krumholz2004} compute the sink accretion rates based on the theory by Bondi, Hoyle and Littleton \citep{Hoyle1939, Bondi1952}. The Bondi-Hoyle radius is
\begin{equation}\label{BH_radius}
	r_{\text{BH}}=\frac{GM_{\star}}{(v_{\infty}^2+c_{\infty}^2)}
\end{equation} 
and the corresponding accretion rate is given by
%\begin{equation}\label{BH_rate}
%	\dot{M}_{\text{BH}}=4\pi \rho_{\infty} G^2M^2 \Bigg[\frac{\lambda^2 c_{\infty}^2+v_{\infty}^2}{\big( c_{\infty}^2+v_{\infty}^2\big)^4}\Bigg]^{1/2},
%\end{equation} 
\begin{equation}\label{BH_rate}
	\dot{M}_{\text{BH}}=4\pi \rho_{\infty} r_{\text{BH}}^2 \sqrt{\lambda^2c_{\infty}^2+v_{\infty}^2 },
\end{equation} 
where $M_{\star}$ is the mass of the star and $v_{\infty}$, $c_{\infty}$, $\rho_{\infty}$ are the velocity of the gas relative to the star, the sound speed and the density far from the star relatively. The parameter $\lambda$ depends on the equation of state, $ \exp(3/2)/4\approx1.12$ is the correct value for isothermal gas. When computing the sink accretion rate, we replace $M_{\star}$ by the sum of the sink mass and the gas mass inside the sink radius to increase the accretion rate of very low mass sinks. Using the recipe given by Krumholz, we choose $v_{\infty}$, $c_{\infty}$ to be the values at the sink location and we extrapolate from the weighted mean density inside the sink accretion radius $\overline{\rho}$ to 
\begin{equation}\label{rho_inf}
	\rho_{\infty}=\frac{\overline \rho}{\alpha(\overline r/r_{\text{BH}})}
\end{equation}	
with $\alpha(x) \equiv \rho(x)/\rho_{\infty}$ being the density profile that arises from the transsonic solution of the spherical Bondi problem as a function of the dimensionless radius $x \equiv r/r_{\text{BH}}$. The radius $\overline r$ corresponding to the density $\overline \rho$ is chosen to match expected results. To average the density inside the sink radius and to smoothen accretion when the sink particle is moving through the grid, we use the same kernel function as \cite{Krumholz2004} which assigns every cell inside the accretion zone a weight 
\begin{equation}\label{kernel}
	w\propto \exp(-r^2/r_{k}^2).
\end{equation}	
Note that in contrast to the description given by \cite{Krumholz2004}, we simply fix $\overline r$ in Equation \ref{rho_inf} as well as the kernel size $r_k$ to half the accretion radius. In the presence of rotational flows around the sink, the Bondi-Hoyle accretion rate is an overestimation of the effective accretion rate. We use trick by \cite{Krumholz2004} to reduce the accretion rate: A cell inside the accretion radius is divided into $8^3$ little sub-cubes. Using the specific energy and the specific angular momentum of the gas, the `closest approach' of each cube to the sink particle is estimated assuming ballistic trajectories.The number of cubes that will not make it closer to the sink than $0.25 \Delta x$ is counted and the Bondi-Hoyle accretion rate is reduced by the corresponding factor.

%%%%%%%%%%%%%%%%%%%%%%%%
%FLUX ACCRETION
%%%%%%%%%%%%%%%%%%%%%%%%
\subsection{Flux Accretion (FA)}\label{flux accretion}
In this accretion method we set the accretion rate equal to the mass flux rate into the sink accretion zone. \cite{Gong2013} first describe this using the fluxes at the cell boundaries returned by the Riemann solver. Since these fluxes are relative to the grid they need to be corrected for the sink motion when a sink particle moves through a density gradient. We therefore take a slightly different approach and compute the mass flux into the accretion zone $\Omega_{\text{acc}}$ using Gauss' divergence theorem,
\begin{equation}\label{flux_acc_rate}
	\dot{M}_{\text{flux}}=-\int_{\Omega_{\text{acc}}} \text{div}\big( \rho(\boldsymbol v - \boldsymbol v_{\text{sink}}) \big) \mathrm dV.
\end{equation}
As we do not allow for negative accretion rates, the gas mass inside the accretion zone can only decrease. To keep the gas density inside the accretion zone close to the sink threshold density in the long term, we correct this mass flux rate by a small factor and use the following `flux accretion rate' 
\begin{equation}\label{acc_rate}
	\dot{M}_{\text{FA}}=\Bigg[1+ 0.1\lg\bigg(\frac{\overline{\rho} }{ \rho_{\text{sink}}}\bigg)\Bigg] \dot{M}_{\text{flux}},
\end{equation}
where $\overline{\rho}$ is the mean gas density inside the accretion zone and  $\rho_{\text{sink}}$ is the user-defined sink threshold. We compute the gas mass that is removed from a cell $\Delta m_i$ in the accretion zone in a mass weighted fashion,
\begin{equation}\label{acc_mass_rate}
	\Delta m_i= \left\{
		\begin{array}{rrl}
			\Delta t \frac{\dot M_{\text{FA}}}{n_{\text{cells}}}  \frac{\rho_i}{\overline{\rho}} & \text{if }&  \dot M_{\text{FA}} \geq 0,\\
   			0 & \text{if }& \dot M_{\text{FA} }< 0,
    		\end{array} 
	\right. 
\end{equation}
where $n_{\text{cells}}$ is the number of cells in the accretion zone.
Since in FA accretion we remove gas from the individual cells in a mass weighted fashion, the gas inside each cell is reduced by the same factor. We make use of this fact to define a new time step criterion to ensure that no cell is emptied completely rather than artificially capping accretion. We compute the total available gas mass inside the accretion zone $M_\text{gas}$ and require
\begin{equation}\label{dt_accretion}
	\Delta t_{\text{acc}} < C \frac{M_\text{gas}}{\dot{M}_{\text{FA}}},
\end{equation}
where we set $C=0.25$ as a default. Using this time step constraint makes sure that not more than 75 per cent of the gas is removed from one cell within a single time step.

%%%%%%%%%%%%%%%%%%%%%%%%
%NO L_ACCRETION
%%%%%%%%%%%%%%%%%%%%%%%%
\subsection{`No-L' Accretion}\label{nol-accretion}
When sink particles accrete gas they remove angular momentum from the simulation. A sink represents a collapsed object which is much smaller then the grid spacing. It is therefore unphysical to simply assign the accreted angular momentum to physical object the sink represents since it would very quickly be spinning at unrealistically high rates. The sink particle therefore acts as a sink not only for the mass, but also for angular momentum. This facilitates accretion from disk-like structures by removing the necessity to transport angular momentum outwards. This was highlighted and found to be important in SPH simulations by \cite{Hubber2013}. They solve this problem by feeding back to the gas the angular momentum that has been accreted previously. We use an approach described by \cite{Krumholz2004}: We decompose the momentum in the motion of the gas relative to the sink into a radial and a tangential part. While the radial part of the momentum is transferred to the sink, the tangential part is assigned to the remaining gas. This corresponds to an acceleration of the remaining gas in the tangential direction since the momentum in the tangential motion remains constant while the gas mass decreases. We keep this `no-L accretion' optional for all accretion schemes. Note that this method does only work if a sink is accreting directly from the gas. In the case where sink formation and subsequent merging work as an accretion mechanism, this technique fails as angular momentum is removed whenever sinks merge.

%%%%%%%%%%%%%%%%%%%%%%%%
%%%%%%%%%%%%%%%%%%%%%%%%
%TESTS
%%%%%%%%%%%%%%%%%%%%%%%%
%%%%%%%%%%%%%%%%%%%%%%%%
\section{Tests}\label{tests}
In this section we report the tests that we have performed using different sink particle implementations. We describe tests on sink formation, sink merging and accretion onto sinks in this order. In the Appendix we discuss two small test cases that concern the integration of the sink trajectories. We try to separate those tests as far as possible which means for example, that when comparing different methods for sink formation, all codes use the same accretion recipe\footnote{This is not always possible, especially since sink formation and merging as it is described by \cite{Krumholz2004} blurs the line between sink formation and accretion.}. 
We test the creation of sink particles using a Boss \& Bodenheimer test (BB test, \cite{Boss1979}) and fragmentation in turbulent molecular gas. We compare three different algorithms for sink formation: a cell-based, for which we use the acronym CELL, a peak-based (acronym PEAK) and a clump-based (acronym CLUMP) strategy. 
\begin{enumerate}
\item In the CELL approach, a sink is formed in every cell that crosses the sink formation threshold $\rho_{\text{sink}}$. Thereby, the gas exceeding the threshold is immediately absorbed by the sink. Sinks are merged using the FOF technique where we have chosen the accretion radius as linking length. 
\item The PEAK strategy discretizes the computational domain by considering every local density peak above $\rho_{\text{sink}}$ for sink formation. A sphere with the size of the accretion radius is defined around the density peak and used as integration domain to compute contraction rates, and energies. The gas inside such a sphere must pass the proximity check, Jeans instability check, bound state check, collapse check and potential minimum check to trigger sink formation. 
\item Our new sink formation algorithm is denoted as CLUMP approach. It allows sinks to be formed only at the density peaks above $\rho_{\text{sink}}$ of clumps having a high enough peak-to-saddle ratio. The gas surrounding those peaks is then subjected to the collapse check, proximity check and the virial check. 
\end{enumerate}
See Section \ref{sink_creation} for a more detailed description of the different checks mentioned above. The turbulent setup is used to compare sink merging.
For sink accretion we consider two test cases, spherical Bondi accretion and accretion from a disk. Those two test cases are applied to the different accretion schemes described in Section \ref{sink_accretion}. We compare Bondi-Hoyle accretion (BH), flux accretion (FA), threshold accretion(TA) and threshold accretion with a threshold reduced by a factor of 10 (TA-low).

%%%%%%%%%%%%%%%%%%%%%%%%
%BOSS BODENHEIMER TESTS
%%%%%%%%%%%%%%%%%%%%%%%%
\subsection{Boss \& Bodenheimer Test \label{bb}}

We performed a series of tests where we followed the collapse and fragmentation of a rotating core, known as the Boss \& Bodenheimer (BB) test. This test consists of a gas sphere in solid body rotation which is seeded with a $m=2$ density perturbation. The sphere collapses into one or more fragments, depending on the parameters used in the setup, and most importantly, on the sink particle algorithm used. BB tests have been used extensively by many authors to test fragmentation in hydrodynamical codes in general and perform resolution studies and code comparisons \citep[e.g.,][]{Boss1979,Truelove1997,Bate1997,Commercon2008}. We choose the same initial conditions as \cite{Federrath2010} when they tested their sink particle algorithm. The parameters of the setup are shown in Table \ref{table:bb_params}. The threshold density for sink formation is chosen as the density above which the local Jeans length is not resolved by 4 cells anymore. A cell is refined when the local Jeans length is less than $4\Delta x$. We use the flux accretion scheme for this test and the sink accelerations are computed using direct force summation.

\begin{table}
	\caption{Simulation parameters for the Boss \& Bodenheimer test.}
	\label{table:bb_params}
	\renewcommand{\arraystretch}{1.2}
	\begin{tabularx}{\columnwidth}{ l r@{ = } l}
	\toprule	
		Radius		 	 			&$R$					&$\SI{5.0e16}{\cm}\approx \SI{3300}{au}$ \\
		Mass		 	 			&$M$					&$\SI{1} {\Msun}$ \\
		Average density 	 			&$\rho_0$				&$\SI{3.82e-18}{\gram\per\cm^3}$ \\
		Free-fall time					&$t_{\text{ff}}$				&$\SI{1.075e12}{\second} \approx \SI{34}{\kilo\year} $\\
		Density perturbation 			&$\rho(\phi)$				&$\rho_0 (1+0.1\cos(2\phi))$ \\
		Isothermal sound speed 			&$c_s$					&$\SI[per-mode = symbol]{1.66e4}{\cm\per\second}$ \\
		$E_{\text{therm}}/E_{\text{grav}}$ 	&$\alpha $ 				&0.26\\
		Angular velocity 				&$\Omega$				&$\SI{7.2e-13}{\second^{-1}} $\\
		$E_{\text{rot}}/E_{\text{grav}}$ 		&$\beta $ 					&0.18\\
		\midrule
		Box size 						&$L_{\text{box}}$			&$\SI{2.0e17}{\cm}$\\% \approx \SI{1.3e4}{au}$\\
		Cell size at levelmax 			&$\Delta x_{\mathrm{min}}$	&$\SI{6.5}{au}$ \\
		Sink radius					&$R_{\text{acc}}$			&$4\Delta x_{\mathrm{min}} \approx \SI{26}{au}$ \\			
		Sink density threshold			&$\rho_{\text{sink}}$			&	$	\left\{\begin{array}{lc}	
																		\SI{8.5e-14}{\gram\per\cm^3} &\text{(iso)}\\
																		\SI{5.5e-13}{\gram\per\cm^3} &\text{(poly)}\\
																	\end{array}\right.
																$\\

		\bottomrule
	\end{tabularx}
\end{table}

\begin{figure*}
	\includegraphics[width=\textwidth]{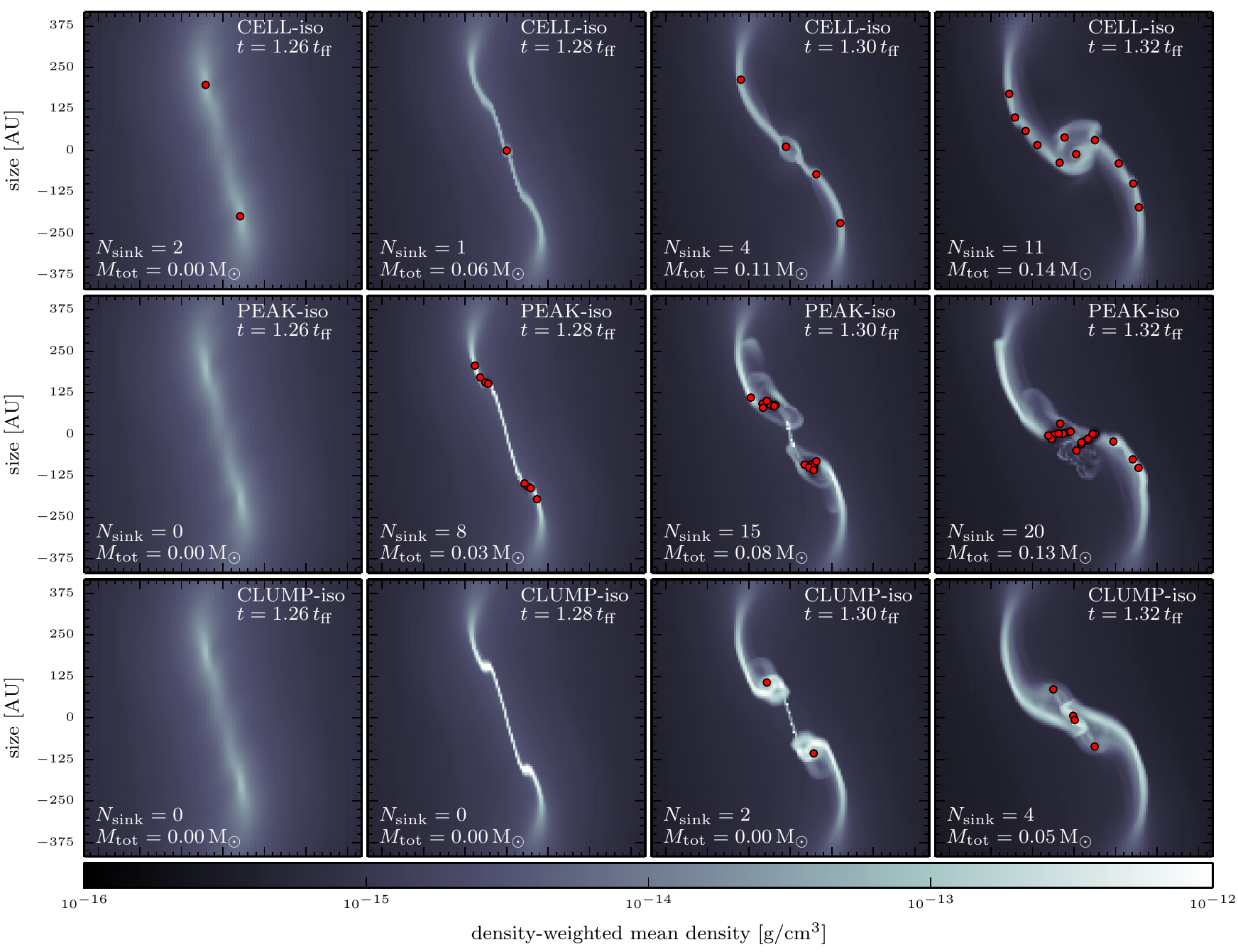}
	\caption{Comparison of different sink formation algorithms on an isothermal Boss \& Bodenheimer test. The time when each snapshot was taken is given in terms of $t_{\text{ff}}\approx 34.1\,  \text{kyr}$. Sink particles are marked with red dots and the size of the dots corresponds to the sink accretion radius. The cell-based algorithm (top row) successfully prevents violation of the Truelove criterion by forming sinks in all cells that cross the density threshold. During the subsequent evolution, constantly ongoing sink formation and merging act as an effective way of accretion and lead to roughly equally spaced sinks along the filament. The peak-based method (middle row) forms 20 sinks from artificial fragments while our new clump-based algorithm (bottom row) allows only 4 of the artificial fragments to trigger formation of a sink during the course of our experiment.}
	\label{bbfigs_iso}
\end{figure*}

\begin{figure*}
	\includegraphics[width=\textwidth]{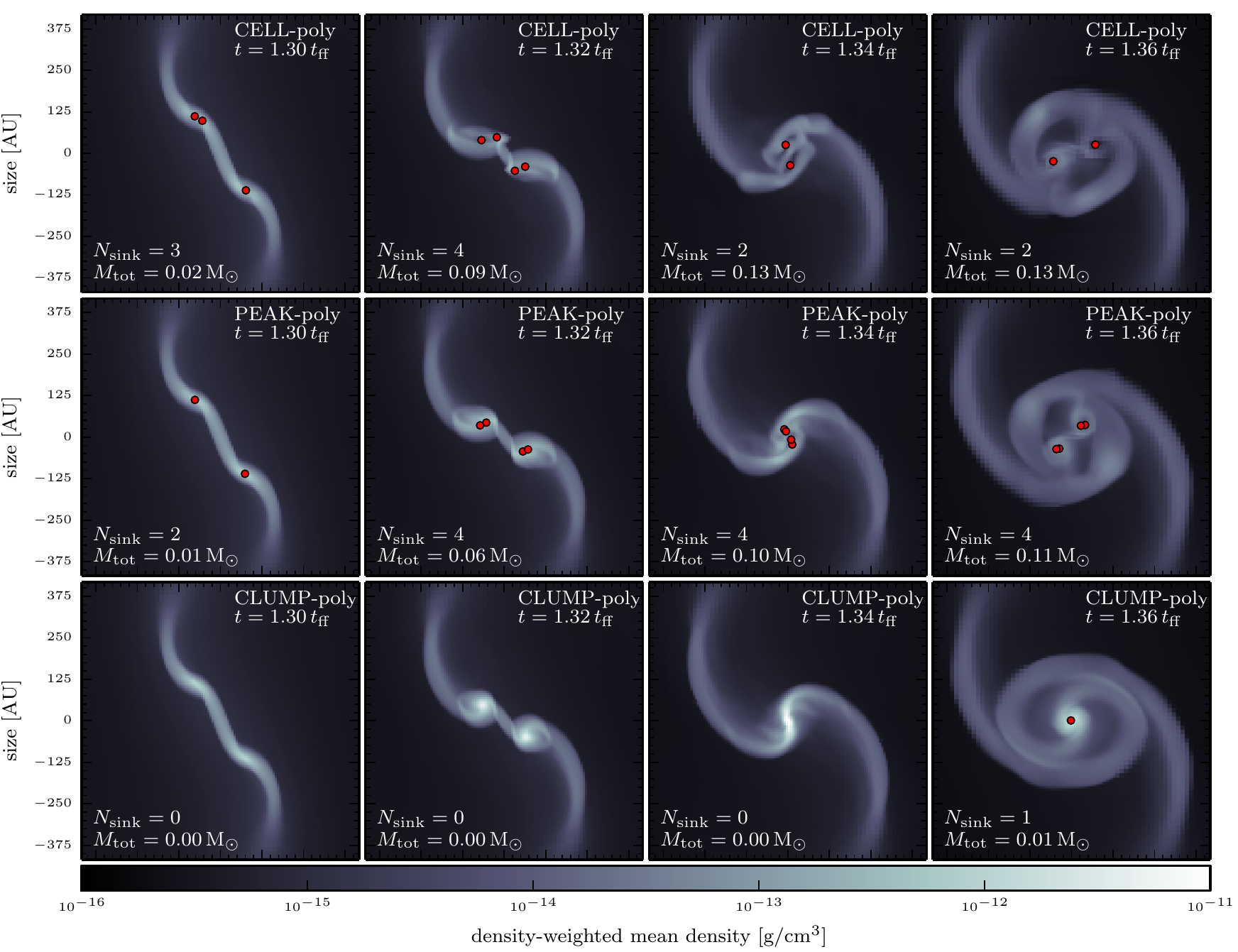}
	\caption{Same as Figure \ref{bbfigs_iso} when using the piecewise polytropic EOS \ref{EOS}. The heating causes the filament to form two distinct fragments. The cell-based method (top row) forms multiple sinks in both fragments that later merge into two sinks forming a binary system. The peak-based algorithm (middle row) triggers formation of a very tight binary inside each fragment. These two binaries orbit each other on a trajectory similar to the one observed in the run where the cell-based sink formation criteria are used. The clump-based method (bottom row) does not allow those two fragments to form a sink as they have too much rotational (the second panel from the left shows that each fragment is in fact a small disk-like structure) and thermal support. Only after the two fragments collide, enough low angular momentum gas is left in the center to form a sink.}
	\label{bbfigs_poly}
\end{figure*}

\subsubsection{Isothermal EOS \label{iso_bb}}
\begin{figure}
	\includegraphics[width=\columnwidth]{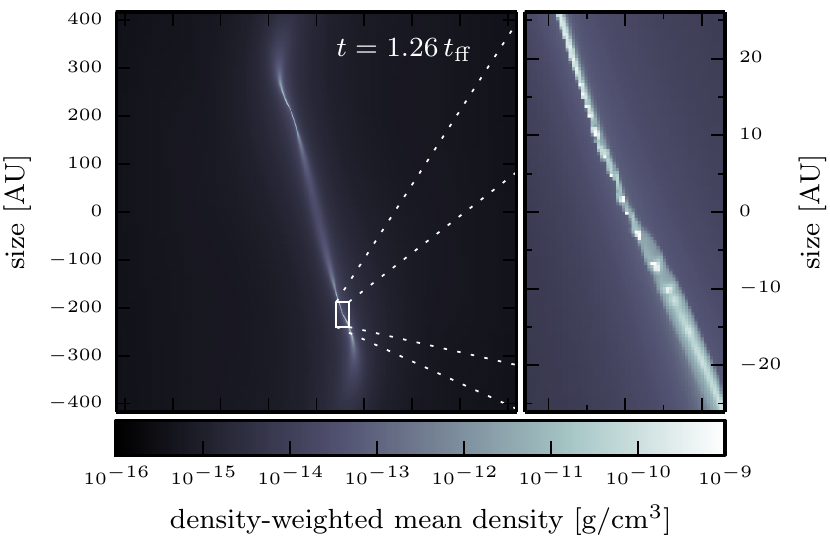}
	\caption{Zoom snapshot of a high resolution isothermal Boss \& Bodenheimer test showing artificial fragmentation.}
	\label{fig:bbhighres}
\end{figure}

For isothermal gas in the absence of magnetic fields, the initial $m=2$ perturbation collapses and forms a filament. No matter what resolution is chosen, this filament will eventually become dense enough to violate the Truelove criterion and fragment artificially \citep{Truelove1997}. This can be observed in Figure \ref{fig:bbhighres} which shows a snapshot for the setup specified in Table \ref{table:bb_params} but including 4 additional levels of refinement, setting the minimum cell size to $\Delta x_{\text{min}}=\SI{0.4}{au}$. Artificial fragmentation is clearly visible.

The filamentary nature of the collapse makes the isothermal BB test a `worst case' scenario for sink formation. Forming sinks in a filament or a sheet will always introduce an artificial length scale which corresponds to the typical sink spacing. However, knowing the behavior of the sink formation algorithm when applied to a collapsing filament is relevant since we know from previous simulations of supersonic turbulence \citep[e.g.,][]{Klessen2004,Heitsch2008,Collins2012} that gas tends to assemble filaments. From observations we know that filaments are ubiquitous in star-forming clouds \citep{Andre2010}.

Figure \ref{bbfigs_iso} shows that our 3 methods for sink formation lead to very different results. The CELL algorithm successfully prevents a violation of the Truelove criterion by immediately absorbing gas that exceeds the density threshold into sinks. The ongoing process of sink creation, accretion and merging results in roughly equally spaced sinks along the filament. The spacing is determined by the resolution dependent sink accretion radius which acts as linking length in the FOF algorithm. As mass is accreted from the continuous 1-dimensional filament onto the discrete number of sinks, the filament is effectively fragmenting on a resolution dependent scale, very similar to the artificial fragmentation in the \cite{Truelove1997} sense. In contrast, the PEAK as well as the CLUMP method do not form a sink until the filament has fragmented artificially. While the PEAK scheme triggers sink formation in almost every artificial fragment, the CLUMP approach is more restrictive and allows only 4 sinks to form before we stop the experiment. The clump finder together with the virial check can prevent most of the artificial fragments from forming a sink. Only those artificial fragments which are dominating the local gravitational field will trigger formation of a sink. Note that sink formation in all 3 cases is still ongoing after the last snapshot shown in Figure \ref{bbfigs_iso}.

\subsubsection{Piecewise polytropic EOS \label{poly_bb}}
Heating the gas is a possible way to prevent the filamentary `catastrophy' described in the last section. We thus repeat the test introducing the same piecewise polytropic equation of state (EOS) as \cite{Federrath2010},
\begin{equation}\label{EOS}	
P= \left\{
    \begin{array}{lll}
     {c_s}^2 \rho 			& \text{if }	& \rho \leq   2.5 \times 10^{-16} \frac{\text{g}} {\text{cm}^3},\\
      \kappa_1 \rho^{1.1}	& \text{if }	& 2.5 \times 10^{-16} \frac{\text{g}} {\text{cm}^3} <   \rho \leq 5.0 \times 10^{-15} \frac{\text{g}} {\text{cm}^3},\\
      \kappa_2 \rho^{4/3}	& \text{if }	&  5.0 \times 10^{-15} \frac{\text{g}} {\text{cm}^3} \le  \rho ,
    \end{array}\right.
\end{equation}
where the values $\kappa_1$ and $\kappa_2$ are chosen such that $P$ is a continuous function of $\rho$. When using this EOS the heating slows down the collapse onto the filament and causes the formation of a well defined fragment at each end of the filament (see Figure \ref{bbfigs_poly}). The CELL run forms and merges sinks in both fragments leading to two sinks forming a binary system. The PEAK run triggers formation of two sinks in each fragment. Note that we do not allow sinks to merge when using the PEAK method for sink formation. When sink merging is turned on, the two sinks inside each fragment merge quickly after the formation of the second sink and the subsequent evolution is very close to the one seen in the CELL run. The CLUMP method identifies the density peak inside each fragment as a possible location for sink formation. Yet both of the fragments fail the virial check due to a combination of rotational and thermal support. Note that although the Truelove criterion is violated and the local Jeans length is not resolved by 4 cells inside the fragments, there is no artificial fragmentation happening. At $t=1.34t_{\text{ff}}$ the two initial fragments undergo a grazing collision leading to ejection of some high angular momentum gas and one fragment in the center which then forms a single sink. 

At this place we want to add a note on the issue of numerical convergence. The isothermal setup is scale free and the fragmentation scale is therefore determined by the artificial fragmentation at the grid scale. More generally, \cite{Martel2006} showed that the fragmentation scale is resolution dependent for isothermal SPH simulations. Consequently, there is no numerical convergence for the isothermal case. The piecewise polytropic case deserves a little more attention. The `knee' in the EOS introduces a physical scale that determines the properties of the resulting fragments \citep{Larson2005}. It seems therefore possible that, once the fragmentation scale is properly resolved, changes in the resolution will not change the results of the numerical experiment anymore. We have thus performed a convergence study on the piecewise polytropic setup where we have increased the sink density threshold according to the numerical resolution. We found that the results for all three sink formation algorithms to be \emph{not} converged in this sense. To understand this behavior one can consider the case of a polytropic index of $5/3$. In this case, the heating is so strong that the increasing pressure will eventually stop the collapse of the fragments, resulting in a stable hydrostatic configuration. Increasing the sink density threshold will therefore at some point prevent sink formation completely. In the case of a polytropic index of $4/3$ there is no stable polytrope \citep{Bonnor1958} and every fragment must collapse eventually. However, we found that by increasing the sink threshold density, one can arbitrarily delay the moment when this threshold is crossed. This is critical as the fragments are in violent dynamical interactions while they are contracting. Delaying the moment of sink formation will therefore alter the results and prevent convergence. It is therefore the physical setup itself which is not converging. One situation where we can imagine convergence in the above sense is the isothermal collapse of a spherical gas configuration as it is probably the case when resolving the second core collapse. Another way to approach the issue of numerical convergence in the presence of sinks is by arguing that the sink density threshold is a physical rather than a numerical parameter and therefore kept fixed as the resolution increases. We do believe that this type of numerical convergence can be achieved for the above setup. Explicitly demonstrating this type of convergence is beyond the scope of this paper.

%%%%%%%%%%%%%%%%%%%%%%%%
%%%%%%%%%%%%%%%%%%%%%%%%
%%%%%%%%%%%%%%%%%%%%%%%%
%TURBCORE
%%%%%%%%%%%%%%%%%%%%%%%%
%%%%%%%%%%%%%%%%%%%%%%%%
%%%%%%%%%%%%%%%%%%%%%%%%
\subsection{Collapse of a Turbulent Molecular Cloud \label{turbcore}}
\begin{table}
	\caption{Physical and numerical parameters for the collapsing molecular cloud test. The two setups are generated using different random number seeds (s1/s2) and slightly different normalizations of the velocity field.}
	\label{table:core_params}
	\renewcommand{\arraystretch}{1.2}
	\begin{tabularx}{\columnwidth}{ l r@{ = } l}
	\toprule
		Radius 		 	 					&$R$					&$\SI{3.0e17}{\cm} \approx \SI{0.01}{\parsec}$\\
		Mass		 	 					&$M$					&$\SI{100} {\Msun}$ \\
		Density		 	 					&$\rho$					&$\SI{1.76e-18} {\gram\per\cm^3}$ \\
		Free-fall time							&$t_{\text{ff}}$				&$\SI{5.0e4}{\year}$ \\
		Mean molecular weight					&$\mu$					&$2.3$ \\
		Temperature	 	 					&$T$					&$\SI{20}{\kelvin}$ \\
		Isothermal sound speed					&$c_s$					&$\SI[per-mode = symbol]{2.68e4} {\cm\per\second}$ \\
		Sound crossing time						&$t_{\text{sound}}$			&$\SI{7.1e5}{\year}$ \\	
		$E_{\text{therm}}/E_{\text{grav}}$			&$\alpha$					&0.04 \\
		Turbulent mach number					&$\mathcal{M}_{\text{rms}}$	&$ 3.65$  / $3.33$\\
		Turbulent crossing time					&$t_{\text{turb}}$			&$\SI{1.9e5}{\year}$ / $\SI{2.1e5}{\year}$ \\	
		$E_{\text{kin}}/E_{\text{grav}}$				&$\beta$					&  $0.18   \,\, /   \, \, 0.15$ \, \\
		$E_{\text{solenoidal}}/E_{\text{compressive}}$	&$\gamma$				&$ \underbrace{1.82}_{s1} \, / \,  \underbrace{1.53}_{s2}$\\

		\midrule
		Box size 					&$L_{\text{box}}$			&$\SI{1.60e18}{\cm}$\\
		max level of refinement		&$l_{\text{max}}$			&13	\\
		min level of refinement		&$l_{\text{min}}$			&8	\\
		Cell size at levelmax			&$\Delta x_{\text{min}}$		&$\SI{13.05}{au}$\\
		Sink accretion radius		&$R_{\text{acc}}$			&$3\Delta x_{\text{min}}$\\
		Sink softening 				&$R_{\text{soft}}$			&$1.5\Delta x_{\text{min}}$\\	
		Sink threshold				&$\rho_{\text{sink}}$			&$\SI{2.46e-14} {\gram\per\cm^3}$ \\
		Mass resolution element		&$m_{\text{res}}$			&$\Delta x_{\text{min}}^3 \rho_{\text{sink}}\approx 10^{-4}\Msol$\\
		\bottomrule
	\end{tabularx}
\end{table}
\begin{figure*}
	\includegraphics[width=\textwidth]{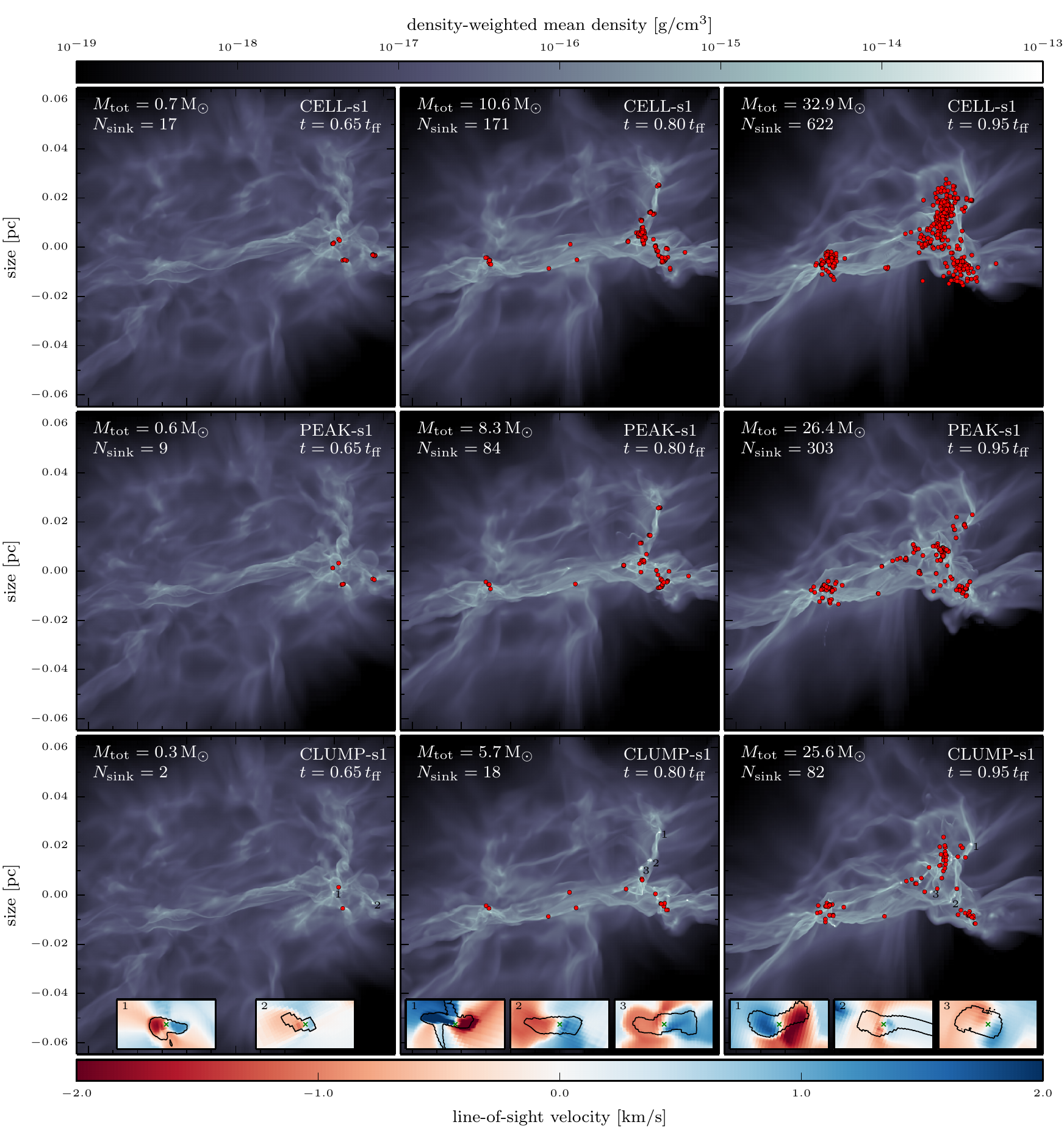}
	\caption{Snapshots comparing the evolution and sink formation of the s1 run for the different sink formation algorithms. The number of sinks and the total mass in sinks is indicated in each snapshot. Sink particles are marked as red dots where the size of the dots is exaggerated in order to be visible and thus not to scale with the rest of the image. The top row shows the results for the cell-based, the middle row for the peak-based and the bottom row for our new clump-based algorithm. The little inlets in the bottom row show enlargements of the most prominent regions that have not yet triggered sink formation by the CLUMP algorithm. The regions are indicated with a little number in the corresponding snapshot. Each inlet covers $\SI{500} {au} \times \SI{250} {au}$ in size and shows a cut plane through the density peak which is oriented along the angular momentum of the gas surrounding the peak. The black line shows the density contour at $\rho_{\text{contour}}=\SI{1.e-14} {\gram\per\cm^3}$ and the color indicates the velocity component perpendicular to that plane. The inlets thus show that the densest sink-less regions are little disks that have considerable rotational support. Therefore these disks fail the virial check and form no sink as they are not undergoing gravitational collapse. }
	\label{3by3_inlets}
\end{figure*}

\begin{figure*}
	\includegraphics[width=\textwidth]{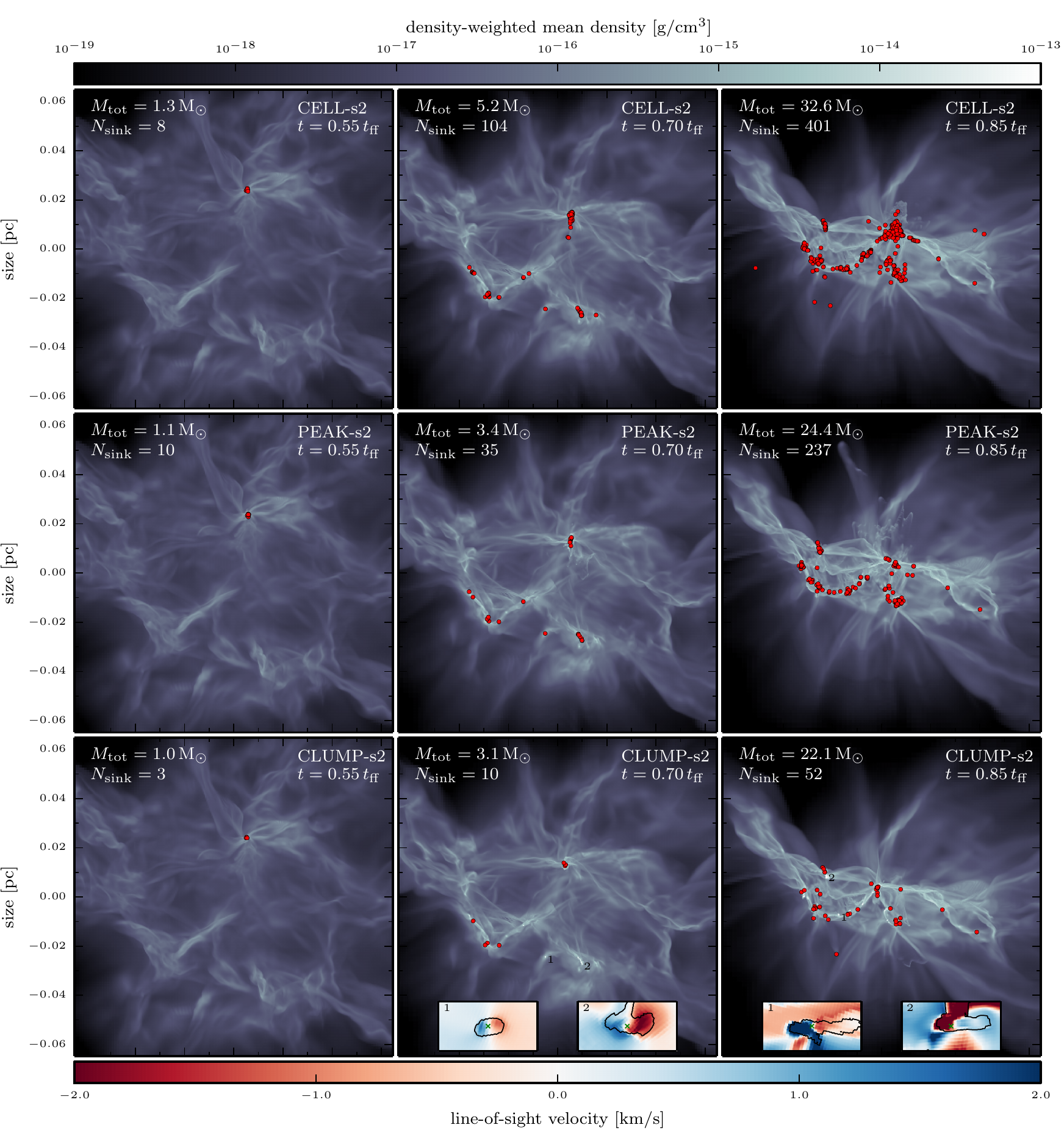}
	\caption{The same as Figure \ref{3by3_inlets} but for the s2 setup. }
	\label{3by3_inlets_456}
\end{figure*}

\begin{figure}
	\includegraphics[width=\columnwidth]{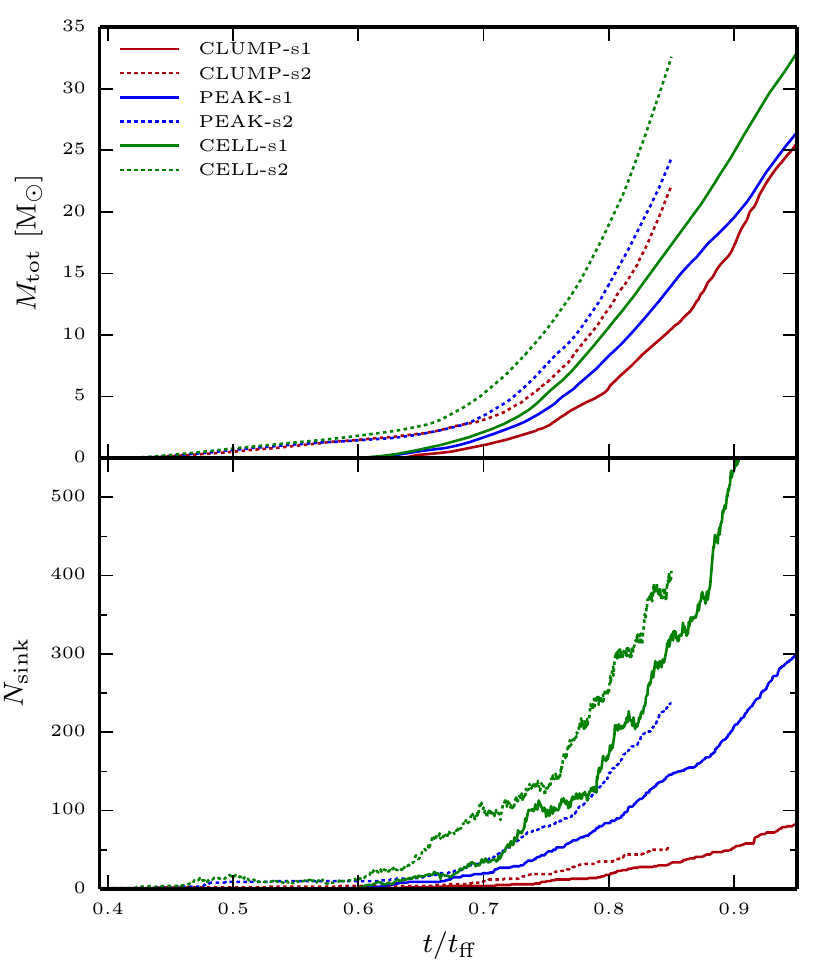}
	\caption{Temporal evolution of the number of sinks $N_{\text{sink}}$ and the total mass in sinks $M_{\text{tot}}$ for the six runs.  }
	\label{turb_sink_t}
\end{figure}

\begin{figure}
	\includegraphics[width=\columnwidth]{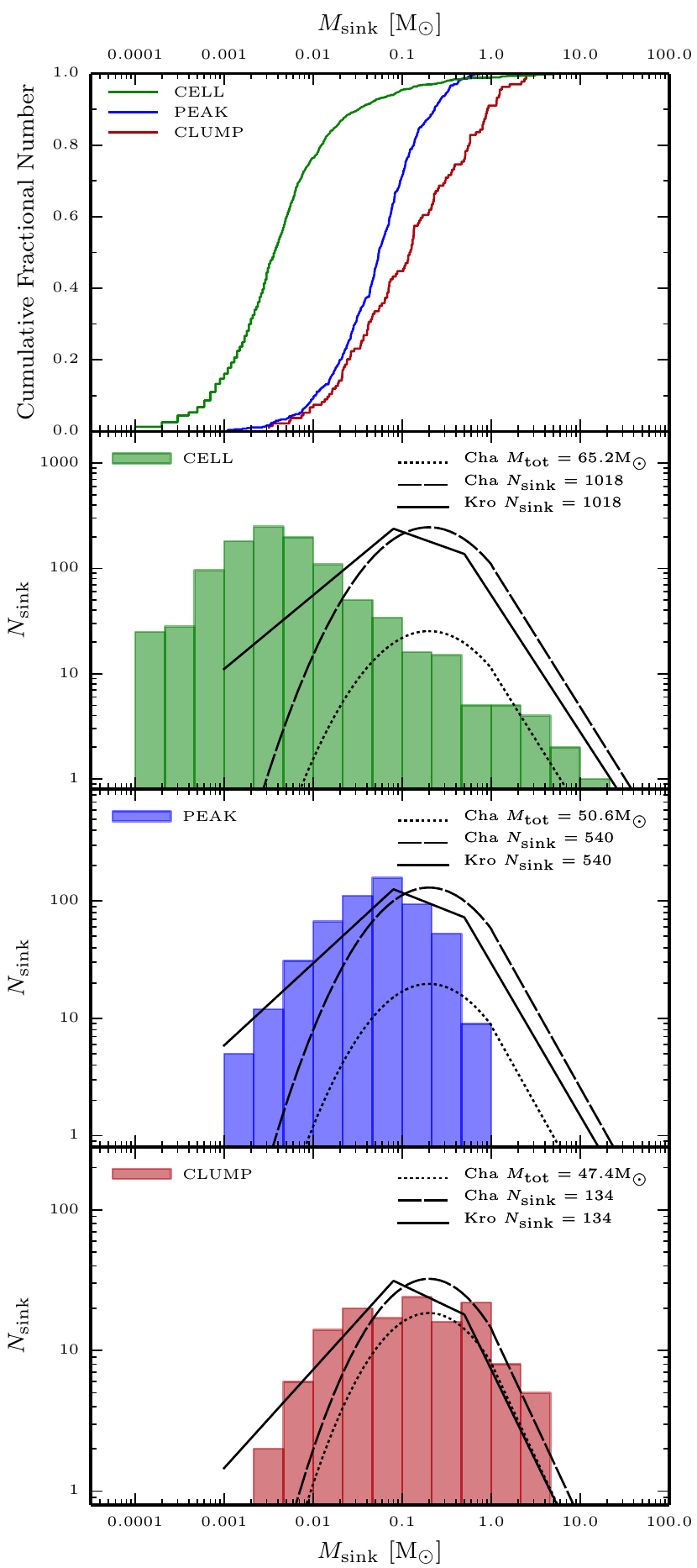}
	\caption{Joint (resulting from the s1- and s2-setup) sink mass distributions for the different sink formation criteria. The top panel shows the cumulative fractional number for each of the three mass distributions in one plot. Below we plot the individual mass histograms together with the \protect\cite{Chabrier2005} IMF normalized to the total mass in sinks and to the number of sinks respectively as well as the \protect\cite{Kroupa2001} IMF normalized to the number of sinks. A Kolmogorov-Smirnov test confirms that all sink mass distributions are different.}
	\label{massfunctions}
\end{figure}

\begin{figure}
	\includegraphics[width=\columnwidth]{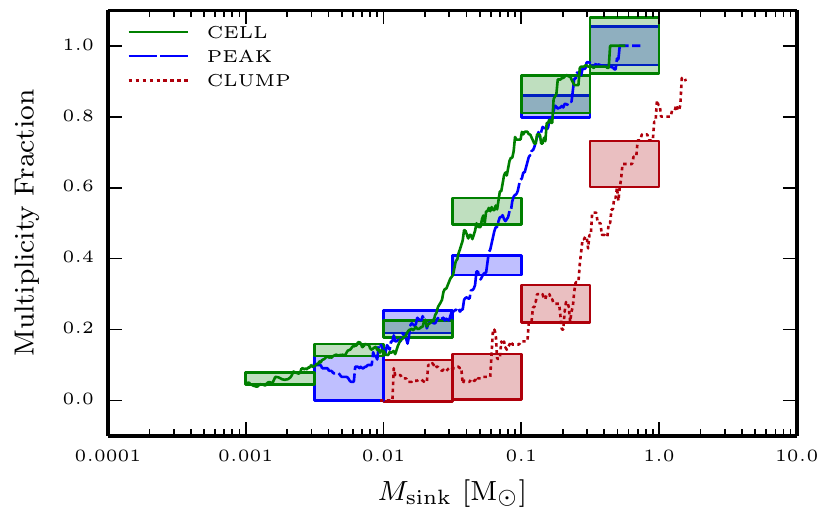}
	\caption{Multiplicity fraction as a function of primary mass. The width of the boxes corresponds to the primary mass bins and the height gives the $\pm 1 \sigma$ range. Only those mass bins with at least 10 objects are considered. The continuous lines show the corresponding boxcar-averages. }
	\label{multifrac}
\end{figure}

Sink particles are an essential ingredient of simulations that model the formation of a star cluster inside molecular gas \citep[e.g.,][]{Girichidis2011,Krumholz2012,Bate2012}. We use such a scenario to compare the different sink formation methods. We use two setups as similar as possible to the top hat runs in \cite{Girichidis2011}. An isothermal, initially spherical gas configuration is seeded with turbulent motions that decay, allowing the cloud\footnote{An object of $100\, \Msol$ would usually by considered a `clump' inside a molecular cloud rather than a `cloud' itself. We label it as `cloud' because we use the word clump already for a much smaller structure in the context of sink formation.}
 to collapse and fragment. Some physical and numerical parameters for this test are summarized in Table \ref{table:core_params}. The velocity field is modeled by Burgers turbulence ($P(k)\propto k^{-4}$) which is in agreement with measured size-linewidth relations in molecular clouds \citep{Larson1981, Heyer2009}. We use mixed turbulence which means that the initial velocity field contains solenoidal (divergence free) as well as compressive (curl free) modes. The sink formation threshold $\rho_{\text{sink}}$ is chosen such that the Jeans length at this density is resolved by exactly 4 cells at the finest level. We use a mass based variant of the Jeans refinement criterion which guarantees that the \emph{smallest} Jeans mass in the calculation is resolved by a fixed number of cells throughout the whole calculation. We therefore compute the mass in one cell at the maximum density $\rho_{\text{sink}}$ and use this as a mass resolution element. During the calculation, a cell is refined as soon as its gas mass exceeds the mass resolution element. This leads to a roughly constant number of cells ($\gtrsim 10^6$) resolving the collapsing cloud throughout the whole calculation and prevents the code from de-refining to low levels early in the calculation when the Jeans length is still large.

We applied each of the 3 methods for sink formation (CELL/PEAK/CLUMP) to both setups (s1/s2) leading to a total of 6 runs. The s1 runs are stopped at $t=0.95t_{\text{ff}}$ and the s2 runs at $t=0.85t_{\text{ff}}$. By this time a total mass of $> \SI{20} {\Msun}$ has assembled in sinks in each run corresponding to a star formation efficiency of $> 20$ per cent. We use the same accretion scheme (FA, no-L accretion, see Section \ref{sink_accretion}) for all runs. The sinks accelerations are computed as direct sums (see Section \ref{sink_integration}). 
Figures \ref{3by3_inlets} and \ref{3by3_inlets_456} show the temporal evolution of the cloud and the sinks formed by each of the three sink formation algorithms for the two setups. The large scale evolution of the cloud is barely affected by the differences in the sink algorithms but the small scale structure of the gas as well as number and properties of the sinks formed do differ. The most remarkable property seen in these snapshots is the high density regions marked with little numbers in the bottom row. These regions are relatively dense and massive but the CLUMP algorithm has not formed a sink at the time the snapshot was taken. Closer inspection of those regions yields strong vorticity in the velocity field and an internal kinetic energy which is $\approx \frac 1 2 E_{\text{grav}}$ which causes the virial check to prevent sink formation. The fact that those dense regions are actually little disks is shown in the inlets in the bottom row of Figure \ref{3by3_inlets}.

In Figure \ref{turb_sink_t} we plot the number of sinks and the total mass in sinks as a function of time for the 6 runs. It is apparent that the total mass in sinks mainly depends on the initial conditions while the details of the sink formation algorithm have a strong effect on the number of sinks formed. Table \ref{table:sink_stats} contains some statistical properties of the sink particle distribution at the end of each run.

\begin{table*}
	\caption{Statistical properties of the sinks formed collapsing turbulent gas using different sink formation algorithms. We show the total number of sinks, the total mass in sinks, the average mass, the median mass and the mass of the heaviest sink at the end of each run. The last column contains the width of the sink mass distribution in log-space.}	
	\label{table:sink_stats}	
	\renewcommand{\arraystretch}{1.2}
	\begin{tabularx}{1.0\textwidth}{ L{0.15\textwidth} L{0.05\textwidth} L{0.05\textwidth} C{0.1\textwidth} C{0.1\textwidth} C{0.1\textwidth} C{0.1\textwidth} C{0.15\textwidth} }
	\toprule
				&IC& $N_{\text{tot}}$		&$M_{\text{tot}} \,[\Msol]$ 		&$\overline M \,[\Msol]$	&$\tilde M \,[\Msol]$	 &$M_{\text{max}} \,[\Msol]$	&$\sigma \Big (\log_{10} (M / \Msol ) \Big)$\\	
	\midrule												
	CELL		&s1	&620	&32.7			&0.053	&0.0042	&9.77		&0.69\\
	CELL		&s2	&398	&32.4			&0.082	&0.0030	&11.67		&0.76\\
	PEAK		&s1	&303	&26.3			&0.087	&0.053	&0.78		&0.51\\
	PEAK		&s2	&237	&24.3			&0.10 	&0.060	&0.75		&0.50\\
	CLUMP		&s1	&82		&25.4			&0.31	&0.080 	&2.65		&0.71\\
	CLUMP		&s2	&52		&22.0			&0.42	&0.18	&2.49	 	&0.63\\																
	\bottomrule			
	\end{tabularx}\\
	\smallskip
\end{table*}

While all sink creation methods agree in the fact that the s1-run forms $\approx 1.5$ times as many sinks as the s2-run, the number of sinks formed and therefore the average sink mass strongly differ. Considering the results of both setups together, the CLUMP algorithm reduces the number of sinks by 87 per cent when compared to the CELL algorithm and by 75 per cent when compared to the PEAK strategy. In Figure \ref{massfunctions} we analyze the joint sink mass functions from both setups for each sink formation algorithm. In the top panel we display the cumulative mass functions. Kolmogorov-Smirnov tests yield p-values below $10^{-8}$ for each pair of mass functions which means that the underlying distributions are different. The absence of further checks for sink formation leads to very high number of low mass sinks produced by the CELL algorithm. Furthermore, the aggressive merging strategy increases the accretion rate of already heavy objects which results in a flat high mass tail and one object with a mass $\sim 10  \, \Msol$ formed in each run. The PEAK and the CLUMP runs produce similarly shaped mass distributions which resemble the observed IMF \citep{Kroupa2001,Chabrier2005}. The PEAK distribution is shifted to lower masses and has a somewhat steeper drop-off at high masses compared to the CLUMP distribution. We find a good qualitative agreement between the PEAK results and the top-hat results obtained by \cite{Girichidis2011} for equivalent setups and a very similar sink formation algorithm. The surprisingly good agreement between our new sink mass function and  the observed IMF (see bottom panel of Figure \ref{massfunctions}) must be seen (at least partially) as a coincidence. The rather low sink formation density threshold of $\SI{2.46e-14} {\gram\per\cm^3}$ and the warm temperature of $20\,\mathrm{K}$ that we have adopted in the numerical experiment both lead to a high minimum Jeans mass which increases the characteristic mass of the produced sinks. These rather arbitrary choices are unavoidable, because we do not model the effect of radiative feedback in setting up the characteristic star particle mass \citep[e.g.,][]{Krumholz2012}. 

In Figure \ref{multifrac} we compare the multiplicity fractions at the end of the simulations. As for the mass functions, we add the results from the s1 and the corresponding s2-run. We adopt the following definition \citep{Hubber2005} of the multiplicity fraction
\begin{equation}\label{multifrac_def}
mf=\frac{B+T+Q}{S+B+T+Q},
\end{equation}
where S is the number of single objects and B,T,Q are the number of binary, triple and quadruple \emph{systems} respectively that have a primary mass in a given range. We follow the algorithm described by \cite{Bate2009} to group the sinks into gravitationally bound systems. Despite the relatively high uncertainty in our results due to the low number of objects per mass bin, one can safely conclude that for the chosen setup a sink with a mass in the range $[0.1 \, \Msol,1.0 \, \Msol]$ has a significantly lower probability to have companions when we use our new sink formation algorithm. We interpret this effect as being due to the correct treatment of tidal forces in our virial check, which hinders the formation of new sinks close to pre-existing ones.

\subsubsection{Sink merging comparison \label{merge_turbcore}}

\begin{table*}
	\caption{Results of the sink merging comparison runs. $N_{+}$ stands for the number of sink formation events, $N_{-}$ for the number of mergers. Furthermore we display the total number of sinks, the total mass in sinks, the average mass, the median massn the mass of the heaviest sink and the width of the sink mass distribution in log-space at the end of each run.}
	\label{table:merging_stats}
	\renewcommand{\arraystretch}{1.2}
	\begin{tabularx}{1.0\textwidth}{ L{0.14\textwidth} L{0.02\textwidth} L{0.04\textwidth} L{0.04\textwidth} L{0.04\textwidth} C{0.09\textwidth} C{0.09\textwidth} C{0.09\textwidth} C{0.09\textwidth} C{0.14\textwidth} }
	\toprule
	
									&IC	&$N_{\text{+}}$	&$N_{-}$	&$N_{\text{tot}}$	&$M_{\text{tot}} \, [\Msol]$ 	&$\overline M \,[\Msol]$	&$\tilde M \,[\Msol]$ 	&$M_{\text{max}} \, [\Msol]$	&$\sigma \Big (\log_{10} (M / \Msol ) \Big)$ \\
		\midrule															
		nomerge						&s1	&82		&0			&82				&25.4		&0.31		&0.08	&2.65			&0.71\\
		nomerge						&s2	&52		&0			&52				&25.2		&0.42		&0.16	&2.49			&0.63\\
		allmerge						&s1	&102	&52			&50				&22.8		&0.46		&0.07	&8.90			&0.77\\
		allmerge						&s2	&51		&23			&28 				&17.0		&0.61		&0.06	&7.30			&1.00\\
		$t_{\text{merge}}=5000\, \text{yr}$	&s1	&106	&54			&52				&24.9		&0.49		&0.07	&5.11			&0.86\\
		$t_{\text{merge}}=5000\, \text{yr}$	&s2	&71		&37			&34			 	&20.1		&0.59		&0.11	&3.72			&0.83\\
		$t_{\text{merge}}=1000\, \text{yr}$	&s1	&106	&37			&69			 	&25.0		&0.36		&0.10	&3.10			&0.78\\
		$t_{\text{merge}}=1000\, \text{yr}$	&s2	&69		&30			&39			 	&22.6		&0.58		&0.16	&5.39			&0.68\\
		$t_{\text{merge}}=500\, \text{yr}$	&s1	&69		&9			&60			 	&23.0		&0.38		&0.10	&2.14			&0.72\\
		$t_{\text{merge}}=500\, \text{yr}$	&s2	&46		&3			&43			 	&22.4		&0.52		&0.20	&3.73			&0.64\\										
		\bottomrule					
	\end{tabularx}\\
\end{table*}

\begin{figure*}
\includegraphics[width=\textwidth]{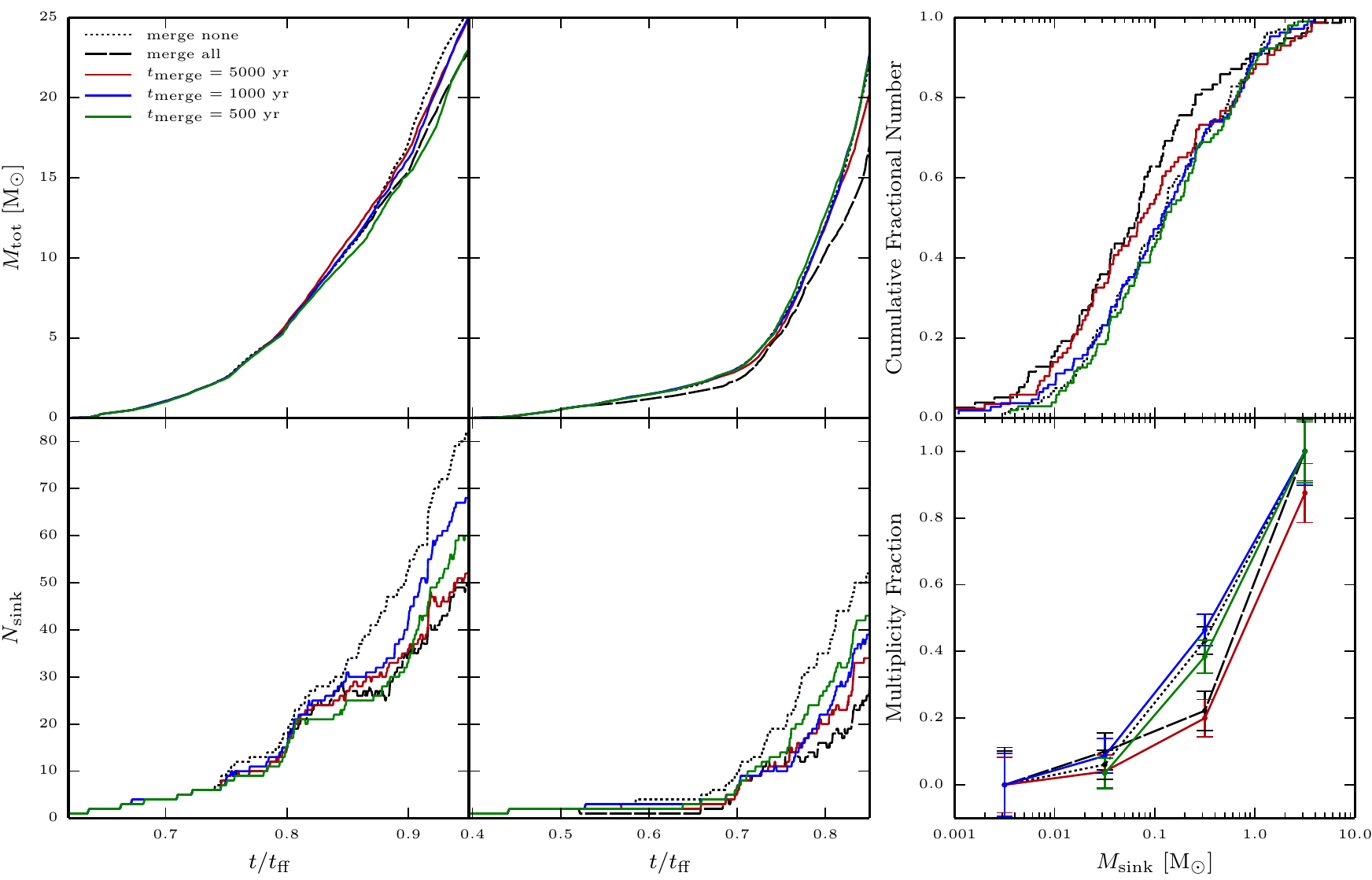}
\caption{Sink merging comparison. The two panels on the left show the number of sinks and the total mass in sinks as a function of time for the s1 setup. The panels in the middle display the corresponding plots for the s2 setup. The upper right panel shows the cumulative sink mass distributions and in the lower right panel we display the multiplicity fraction as a function of primary mass, where each datapoint covers one order of magnitude in primary masses.}
\label{merging_plots}
\end{figure*}

We use the same turbulent core to test the influence of sink merging onto sink formation and accretion. In Section \ref{sink_merging_timescale} we introduced the concept of a merging timescale allowing only young sinks to merge. The same timescale is  used as maximum timescale (or a minimum speed) at which the gas must contract in order to form a sink. The physical motivation for this merging of young sinks is the finite lifetime of the first Larson core \citep{Larson1969, Masunaga1998} of $\sim 1000\, \text{yr}$ during which the sink represents an `fluffy' uncollapsed object. We thus compare the results from the previous section where sink merging is turned off to runs where three merging time scales $t_{\text{merge}}=500\, \text{yr},$  $1000\, \text{yr},$ $5000\, \text{yr}$ and a case where we allow sinks to merge during their entire lifetime corresponding to a infinite merging timescale. Some statistical properties of the sinks formed in each run are listed in the Table \ref{table:merging_stats}. 

In Figure \ref{merging_plots} we plot the temporal evolution of the number of sinks and the mass in sinks together with the cumulative sink mass distribution and the multiplicity fraction as a function of primary mass. Comparing the two limiting cases (no merging, infinite merging lifetime) we find that sink merging reduces the number of sinks by $\approx 40$ per cent. Furthermore we see a strong increase in the mass of the heaviest sink together with slight decrease of the median sink mass, resulting in wider mass distribution. However, the data generated in this test is rather scarce. A Kolmogorov-Smirnov test returns a p-value of 7 per cent when comparing the joint (s1 together with s2 run) distributions resulting from the `nomerge' and the `allmerge' runs. Yet the observed trend fits well with our results for the CELL algorithm in the previous section which merges sinks in a FOF-fashion and produces a wider mass distribution and a couple of very high mass objects too. In the s1-run merging increases the number of sink formation events which suggests that the region close to the site of a merger will often create another sink. However, this seems to be very setup dependent as the s2 run shows a different picture. Here, merging decreases the total mass in sinks through the early formation of a very heavy object that prevents sinks from being formed in its surrounding. 

As one expects, sink merging decreases the number of sinks in multiple systems. The bottom-right panel of Figure \ref{merging_plots} shows a reduction of the multiplicity fraction by $\approx 50$ per cent for primary masses in the range $[0.1 \, \Msol, 1.0 \, \Msol]$ when sinks are merged.

We now take a look at the three cases where we used a finite merging timescale. A merging timescale of $5000 \, \text{yr}$ gives results similar to the `allmerge' case as most of the sink formation is happening within $0.2\, t_{\text{ff}} \approx 10 \, \text{kyr}$. The reduction of the heaviest sink masses shows that the very high mass objects produced by the `allmerge' runs form through late time mergers. The shorter merging timescales lead to results which are more similar to the `nomerge' runs, following a trend for lower maximum mass, slightly higher mean mass and narrower mass distribution for shorter merging timescales. For $t_{\text{merge}}=500 \, \text{yr}$ the usage of the merging timescale to define a minimum contraction rate for sink formation starts to kick in, leading to less formation events than in the case without merging.

We now change our focus to accretion onto sink particles. Accretion can influence formation and merging of sink particles by producing new peaks in the gas density field which might trigger sink formation. It is therefore desirable to have an accretion scheme which produces a smooth transition of the flow variables at the sink accretion boundary.

%%%%%%%%%%%%%%%%%%%%%%%%
%%%%%%%%%%%%%%%%%%%%%%%%
%% ACCRETION TESTS          
%%%%%%%%%%%%%%%%%%%%%%%%
%%%%%%%%%%%%%%%%%%%%%%%%

%%%%%%%%%%%%%%%%%%%%%%%%
% Bondi Sphere
%%%%%%%%%%%%%%%%%%%%%%%%
\subsection{Spherical Bondi Accretion}

\begin{figure}
	\includegraphics[width=\columnwidth]{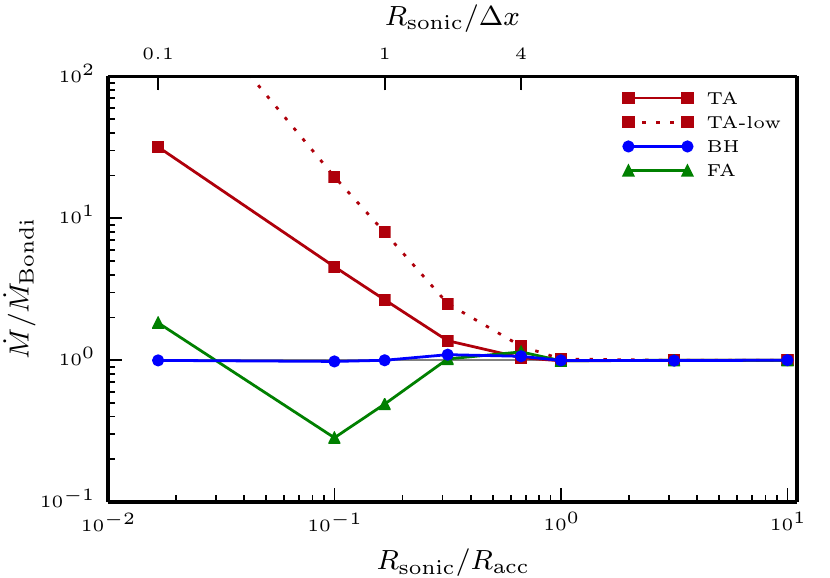}
	\caption{Normalized accretion rates as a function of the sonic radius when simulating the Bondi problem with the different accretion schemes.}
	\label{fig:bondi_ratios}
\end{figure}

There are two setups which are frequently used to test the accretion of sink particles from spherically symmetric gas configurations: The collapse of a singular isothermal sphere first studied by \cite{Shu1977} and Bondi accretion \citep{Bondi1952}. While the collapse of the isothermal sphere is usually well modeled by the codes using sink particles \citep{Krumholz2004, Federrath2010, Gong2013} the Bondi accretion test is a harder challenge as soon as the infall velocity of the gas onto the sink particle is smaller than or of the same order as the sound speed \citep{Krumholz2004, Hubber2013}. We therefore follow the latter two authors and test how well the different accretion methods recover Bondi's transsonic isothermal solution for the accretion of a star at rest relative to the surrounding gas,
\begin{equation}\label{bondi_rate}
	\dot{M}_{\text{Bondi}}=\frac{\pi\exp(3/2)G^2M_{\star}^2\rho_{\infty}}{c_{\infty}^3}, 
\end{equation}
where all the quantities have the same meaning as in Section \ref{Bondi-Hoyle accretion}. We place a spherically symmetric gas ball in a simulation box with an effective resolution of $512^3$ cells. The radius of the ball is 128 cells or 25 per cent of the box. The initial density and velocity field inside the ball are chosen according to the numerical solution of the Bernoulli equation. Outside, the density obtained by the numerical solution is multiplied by $10^{-4}$ and the cells are de-refined by two levels. A sink particle is placed at the center of the box. The accretion radius $R_{\text{acc}}$ is set to 6 cells and the gravitational softening radius $R_{\text{soft}}$ of the sink is 3 cells\footnote{As \cite{Krumholz2004} we find the results to be more accurate when the force at the sink accretion boundary is given by the unsoftened value.}. The sink threshold density is picked according to the numerical solution at the location of the sink boundary. We employ the PM method for computing sink-gas interactions. Using direct force summation instead yields almost identical results. In this test, the total gas mass is negligible compared to the sink mass justifying the assumption of a constant gravitational field.

In the transsonic Bondi solution the sonic radius 
\begin{equation}\label{sonic radius}
	R_{\text{sonic}}=\frac{GM_{\star}}{2c^2}
\end{equation}
separates regions of supersonic gas velocities (inside $R_{\text{sonic}}$) from regions of subsonic flows (outside $R_{\text{sonic}}$). Varying the sink mass therefore sets the ratio $R_{\text{sonic}}/R_{\text{acc}}$ which determines whether the inflow through the sink boundary is subsonic or supersonic. The simulations are stopped at $t_{\text{end}}=4R_{\text{acc}}/c$ which is after the accretion rates have reached constant values but before the rarefaction wave from the boundary enters the scene as $4R_{\text{acc}}\ll R_{\text{sphere}}$. We compare the accretion rates at the end of the simulations to the analytical Bondi rates given by Equation \ref{bondi_rate} and plot it in Figure \ref{fig:bondi_ratios} against the ratio $R_{\text{sonic}}/R_{\text{acc}}$ for the different accretion schemes.

In the supersonic regime  $R_{\text{sonic}}>R_{\text{acc}}$ all the simulated accretion rates differ by less than 1 per cent from the Bondi rate and even when the sonic Radius and the sink radius are the same, all accretion rates are within 5 per cent from the analytic value. Modifications of the density fields within the sink accretion radius are `hidden' from the rest of the simulation domain since no wave can propagate outward from the accretion zone. In the subsonic regime  $R_{\text{sonic}}<R_{\text{acc}}$ the simulated accretion rates differ by many orders of magnitude. Altering the density inside the sink accretion zone now does affect the accretion rate. For instance, an overestimation of the \emph{initial} accretion rate can cause a sharp drop in the density at the boundary of the accretion zone which triggers an outward traveling rarefaction wave and therefore leads to a \emph{permanent} overestimation of the accretion rate. 

Not surprisingly the BH accretion scheme performs best at what it was designed for - solving the Bondi problem. For the BH case we find the biggest deviation from the analytic value when the sonic radius is in between the sink accretion radius and the grid spacing. At $R_{\text{sonic}}/R_{\text{acc}}=0.31$ we overestimate the accretion rate by $9$ per cent. For the regime where $R_{\text{sonic}} \ge R_{\text{acc}}$ or  $R_{\text{sonic}} \le \Delta x$ the errors are smaller than one per cent. This is similar to \cite{Krumholz2004} who find a deviation of $\approx 25$ per cent from the analytic value when the accretion radius is of the same order as the Bondi radius. In the TA case, the accretion rates are very sensitive to the chosen threshold as soon as $R_{\text{sonic}} < R_{\text{acc}}$. Even though we artificially set the sink threshold to the analytic value at the location of the sink accretion boundary, the accretion rate is overestimated. The results are obviously worse for the TA-low case as reducing the density inside the accretion zone reduces the back pressure on the flow outside the accretion zone. On the other side, increasing the threshold by one order of magnitude stops accretion completely in that regime (not plotted). The FA scheme seems to perform acceptably in this test on the first sight, yet it suffers from a different problem: For the runs where  $R_{\text{sonic}}/R_{\text{acc}} \le 0.31$ the accretion rates do not converge during the course of the simulation. Instead of the final value we therefore plot the \emph{average} accretion rates for those data points. The FA scheme lets the sink accrete exactly at the Bondi rate at the beginning of the simulation since the mass flux into the accretion zone is correctly set by the initial conditions. When running the simulation long enough, the accretion rate starts to oscillate with a growing amplitude, temporarily even dropping to zero. We interpret this behavior in the following sense: Stability analysis of the Bondi problem \citep[e.g.,][]{Stellingwerf1978} have shown that only the transsonic solution to the problem is stable. When the resolution is very limited ($R_{\text{sonic}} \lesssim \Delta x$) there is no region where the flow is supersonic and the solution to the problem becomes indistinguishable from solutions without a supersonic region and therefore unstable. The reason that we do not see this instability for the BH and the TA case is that for those schemes, the accretion rates are effectively monotonic functions of the density inside the accretion zone what stabilizes those solutions.

\subsection{Disk Accretion Tests}
Sink particles in simulations of self gravitating turbulent gas accrete most of their mass from the disks that form around them. Since there is no appropriate toy model with analytical solution for this mode of accretion, we have to compare results obtained by using different accretion schemes to each other without knowing the `true' solution. We do this by studying the collapse of a rotating gas sphere which triggers the formation of a sink particle surrounded by an accretion disk. The parameters describing the initial setup together with some simulation parameters are listed in Table \ref{table:disk_params}. Since we do not use radiative feedback in these calculations, we use a piecewise polytropic EOS

\begin{equation}\label{disk_poly_EOS}	
P= \left\{
    \begin{array}{lll}
     {c_s}^2 \rho 			& \text{if }	& \rho \leq   \SI[per-mode = symbol]{1.e-16}{\gram\per\cm^3},\\
      \kappa \rho^{1.4}		&  \text{if }	& \rho \ge   \SI[per-mode = symbol]{1.e-16}{\gram\per\cm^3},
    \end{array}\right.
\end{equation}
to heat the dense gas and prevent the disk from fragmenting into multiple sinks. $\kappa$ is chosen such that $P$ is a continuous function of $\rho$. In this test we use direct force summation for computing sink-gas interactions\footnote{Comparison runs using the PM scheme show similar behavior for the first $\approx \SI{5} {k\year}$, but tend to loose symmetry quickly once the sink is growing massive and therefore dominating the gravitational potential. This causes the sink to leave the center of the disk which considerably changes the results. }.
\begin{table}
	\caption{Simulation parameters for the disk accretion tests.}
	\label{table:disk_params}
	\renewcommand{\arraystretch}{1.2}
	\begin{tabularx}{\columnwidth}{ l l@{ = } l}
		\toprule	
		Sphere radius						&$R$					&$ \SI{2000}{au}$ \\
		Total gas mass		 	 			&$M$					&$\SI{2} {\Msun}$ \\
		Density profile						&$\rho ( r )$ 				&$\displaystyle \frac{\rho_0}{ (r/r_0)^2+1} $\\
										&$\rho_0$				&$\SI[per-mode = symbol]{4.7e-13}{\gram\per\cm^3} $\\
										&$r_0$					&$\SI[per-mode = symbol]{10}{au} $\\
		Isothermal sound speed 				&$c_s$					&$\SI[per-mode = symbol]{1.88e4} {\cm\per\second}$ \\
		$E_{\text{therm}}/E_{\text{grav}}$		&$\alpha$					&$0.06 $\\
		Angular velocity					&$\Omega$				&$\SI[per-mode = symbol]{5.45e-12} {\second^{-1}}$\\
		$E_{\text{rot}}/E_{\text{grav}}$			&$\beta$					&$0.33 $\\
		\midrule

		Box size 							&$L_{\text{box}}$			&$ \SI{32000}{au}$\\
		Cell size at levelmax 				&$\Delta x_{\text{min}}$		&$\SI{7.8}{au}$ \\
		Sink accretion radius				&$R_{\text{acc}}$			&$4 \Delta x_{\text{min}}$ \\
		Sink softening radius				&$R_{\text{soft}}$			&$2 \Delta x_{\text{min}}$ \\

		\bottomrule	
	\end{tabularx}
\end{table}

\begin{figure*}
	\includegraphics[width=\textwidth]{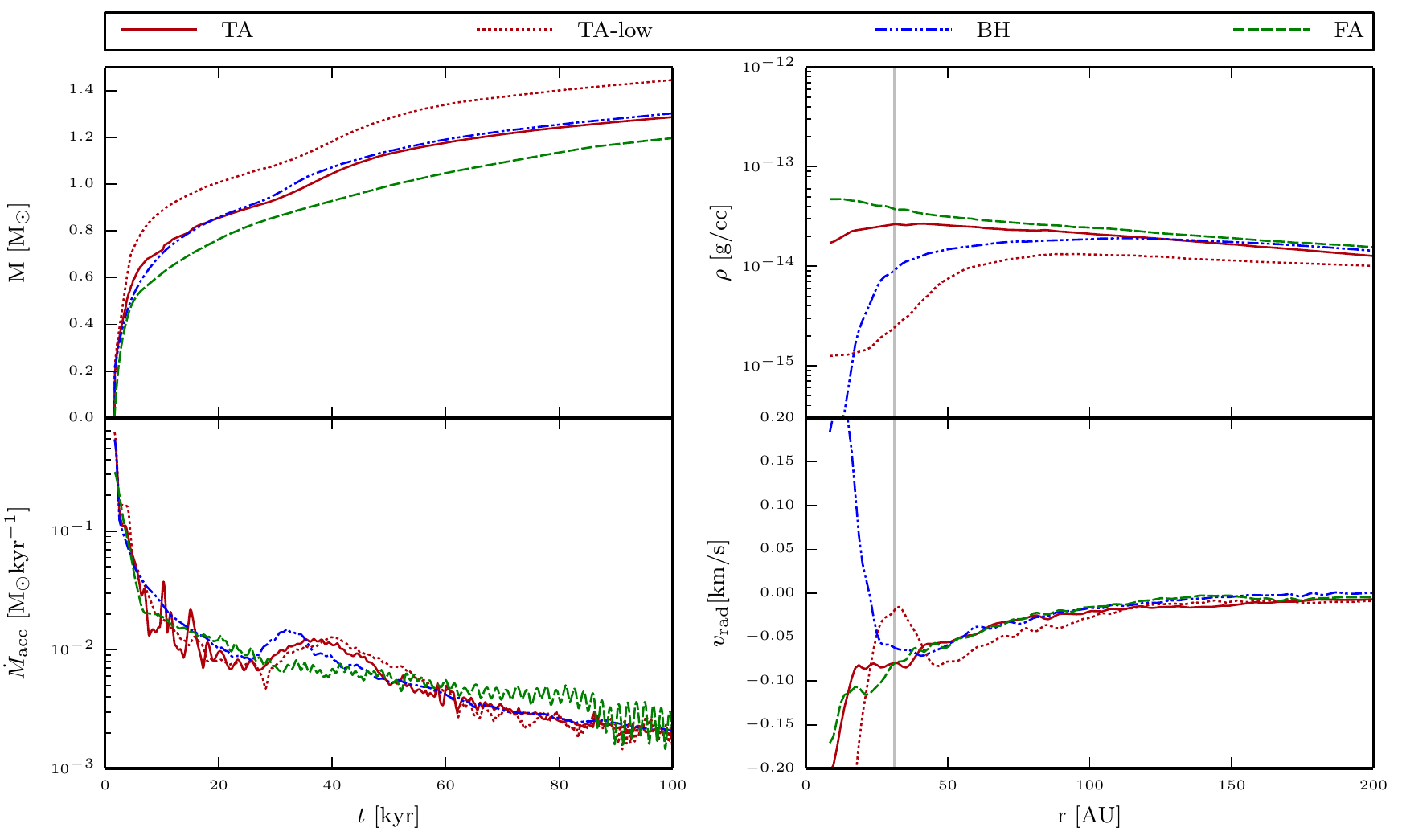}
	\caption{Accretion from a disk onto the central sink particle using different accretion schemes. The two panels on the lefthand side display mass and accretion rate of the central sink particle as a function of time. The panels on the right hand side show the corresponding disk profiles at $t=\SI{50} {k\year}$. The disk profiles are computed in a mass weighted fashion considering all cells within $\pm 39 \, \text{au}$ from the mid plane which corresponds to roughly one scale height in each direction. The vertical gray line at $\approx \SI{30}{au}$ in the profiles indicates the sink accretion radius.}
	\label{acc_rates_profiles}
\end{figure*}

At $ t=\SI{1.61} {k\year}$ a sink forms at the center of the sphere. Very quickly after its formation a marginally unstable disk starts to develop around the sink. At $ t=\SI{5} {k\year}$ the diameter of the disk has reached $\approx \SI{200}{au}$. By this time, the accretion rate has dropped to several $10^{-5} \,\Msol \text{yr}^{-1}$. We let the sink accrete from that disk until we stop the simulation at $t_{\text{end}}=\SI{100} {k\year}$. 
 
The sink masses and accretion rates as a function of time are plotted in Figure \ref{acc_rates_profiles} together with the disk density- and radial velocity profiles at $ t=\SI{50} {k\year}$. The simulated accretion rates differ strongly right after the sink formation. During this phase the disk can efficiently dispose of angular momentum by accreting it into the sink and pressure gradients still play an important role in controlling accretion. A high temporary accretion rate can therefore lead to a high permanent accretion rate. As soon as the disk surrounding the sink is a few times the size of the accretion zone, the accretion rates tend to converge for the different accretion recipes, thereby conserving the differences in their masses that they have obtained in the first $\approx \SI{10} {k\year}$. In this phase we see a self-regulating effect even in the absence of radiative feedback from the sink particle: Lower accretion rates lead to higher densities in the disk which promotes the development of spiral arms. These spiral arms facilitate the re-distribution of angular momentum and therefore increase the accretion rate. 
The upper right panel in Figure \ref{acc_rates_profiles} shows that the density in the center of the disk does depend on the chosen accretion scheme. Because of the subsonic radial velocities (lower right panel) these changes are not restricted to the sink accretion zone but affect the density profile out to several accretion radii. The way accretion is controlled in the BH and TA scheme favors accretion from cells very close to the sink. Together with the centrally peaked accretion kernel used in the BH run, this leads to drop in the density by more than two orders of magnitude. In the TA runs, the depth of this central hole is limited by the accretion threshold. No such hole is produced by the FA run which shows the smoothest transition of the flow into the accretion zone.

\begin{figure}
	\includegraphics[width=\columnwidth]{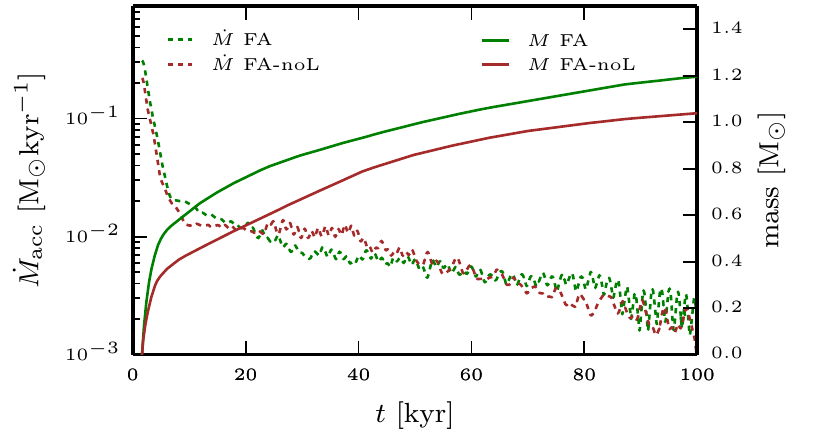}
	\caption{Mass and accretion rate onto the central sink particle when we use no-L accretion compared to the standard flux accretion case.}
	\label{nol_accretion_plot}
\end{figure}

\subsubsection{Accretion of angular momentum}
The above tests have been performed without conserving angular momentum in the gas when accreting onto the sink. We briefly study the effect of what we call no-L accretion (see Section \ref{nol-accretion}) in combination with the FA accretion method using the above setup. \cite{Hubber2013} find that angular momentum feedback from the sink back to the SPH particles considerably lowers the accretion rate during the first $\approx \SI{1} {k\year}$ when following the collapse of a rotating Bonnor-Ebert sphere. We plot our results in Figure \ref{nol_accretion_plot}. Even though AMR as a fundamental difference to SPH does not conserve angular momentum, our results agree well with those found by the authors mentioned above for the early evolution of the sink. During the first $\SI{10} {k\year}$ after the formation of the sink, the average accretion rate is reduced by $30$ per cent  when no-L accretion is used. As soon as the disk is big compared to the accretion radius, the amount of angular momentum that can be advected into the sink particle is small compared to the angular momentum in the disk and the accretion rates for the two runs are very similar. Although not huge, this difference in the early accretion rate might still be enough to affect the probability of a core to fragment into a multiple system rather than a single object.

\section{Conclusions}
We presented a new method for sink particle creation and its implementation in the AMR code \textsc{ramses}. The new method uses a clump finder to identify well-defined density peaks as possible locations for sink formation. We discussed previously introduced tests that are used to examine the gas surrounding a density peak for gravitational collapse and suggested a new criterion based on a virial equilibrium type analysis that fully respects the tidal fields caused by the surrounding mass distribution. We argue that this is more physically motivated than existing criteria. 
We compared the new method to the most frequently used sink creation recipes in simulations of gas undergoing gravitational collapse. Overall, we found our new algorithm to be more restrictive and it triggers less sink formation than other techniques. We showed that none of the sink particle implementations can prevent artificial fragmentation of a filament that formed in an isothermal Boss \& Bodenheimer test. However, our new method is less susceptible to the formation of sinks from those artificial fragments. We simulated the collapse and fragmentation of a small isothermal molecular cloud and found that the number of sink particles formed varies by up to a factor of eight depending on the sink formation algorithm used. The median values of the obtained sink masses differ by up to a factor of 60 and the most massive sinks produced in each run vary by more than one order of magnitude. In the same test, our new algorithm gives rise to a lower probability for sinks to be part of a multiple system than the comparison runs. We do not repeat the analysis for non-isothermal gas, but performing a Boss \& Bodenheimer test using a polytropic EOS suggests that the statistical properties of the sinks formed in non-isothermal turbulent gas will depend on the sink creation routine as well. We therefore conclude that the usage of (different) sink algorithms limits the comparability of results in star cluster formation simulations. Furthermore, great care must be applied when interpreting results that are obtained from such calculations.

We discussed merging of sink particles and describe an intermediate scenario that allows sinks to merge during a certain time-span. We tested sink merging on the turbulent cloud setup. In combination with our new sink creation routine we found that sink merging reduced the number of sinks up to a factor of two when sinks are allowed to merge during their entire lifetime. The information obtained in this test does not allow us to make definite statements about the influence of merging on the sink mass distribution and multiplicity function, but we observed a trend towards a small increase in the width of the mass distribution and a decrease of the multiplicity fraction when merging is allowed. More significantly, sink merging does increase the mass of the most massive sink produced in a calculation. These effects of sink merging can be expected to be even larger when a less restrictive sink formation algorithm is used.

We implemented and compared two schemes for computing sink-sink and sink-gas forces: a PM method and a direct force summation approach. The PM scheme produces surprisingly stable orbits when we let two sinks orbit each other on elliptical trajectories as long as all AMR level boundaries are sufficiently far away. When a sink particle which dominates the local gravitational potential gets close to a level boundary, spurious forces arise that can artificially influence the results. On the other hand, we obtain a speed gain of the order of $\sim (1+\frac{n_{\text{sink}}}{100})$ for a typical setup when using the PM method. A possible way to improve on this situation is to include only the most massive sinks in the direct force summation and to treat the lighter ones using the PM method. 

We have implemented different methods to perform accretion onto sink particles and tested these on two different simulations. The case of spherical Bondi accretion is well modeled by all methods as long as the infall velocity through the sink accretion radius is supersonic. When the accretion is subsonic, only the usage of the Bondi formula for computing the accretion rate will give a correct and stable result in the long run. When accreting from a disk, all accretion schemes yield similar results as soon as the disk radius is larger than a few accretion radii. However, the density and velocity profile of the region close to the sink can be affected considerably. We find that flux-accretion produces the smoothest profiles without any violent changes of the hydrodynamic variables at the sink boundary. Furthermore, flux accretion naturally adapts the accretion rate in the case of a disk with no need for evaluating specific energies on a cell-by-cell basis. These properties lead us to adopt flux accretion as our standard accretion scheme for sink particles in \textsc{ramses}. However, for situations where the the sonic radius of the sink is smaller than the sink accretion radius (as it can be the case for a sink inside an object undergoing Kelvin-Helmholtz contraction) we recommend switching to the Bondi rate. This can be achieved automatically by the simulation code. We implemented so called `no-L accretion' where we leave the angular momentum in the remaining gas which is not accreted. Comparing this to the case where sinks act as sinks for the angular momentum as well yields a considerable reduction of the obtained accretion rate from the disk in the early stage after the sink formation. Once a large disk has formed around the sink, the difference in the accretion rates is negligible.

%\the\textwidth
%\the\columnwidth
%\the\textheight

%normal: \fontname\font\ at \the\fontdimen6\font
%\small{small: \the\fontdimen6\font} \\
%\fontname\font at \the\fontdimen6\font

\section*{Acknowledgements}
The computations leading to this publication have been performed at on the zBox4 and Schroedinger Supercomputers at the University of Zurich and at the Swiss Supercomputing Centre CSCS in Lugano. This work has been supported by the Swiss National Science Foundation SNF under the project `Computational Astrophysics' and the PASC co-design project `Particles and Fields'.

\bibliography{sinks}

%%%%%%%%%%%%%%%%%%%%%%%%
%Integration Test
%%%%%%%%%%%%%%%%%%%%%%%%
\subsection*{Appendix A: Elliptical Orbit Test \label{sink_orbit}}
\begin{figure}
	\includegraphics[width=\columnwidth]{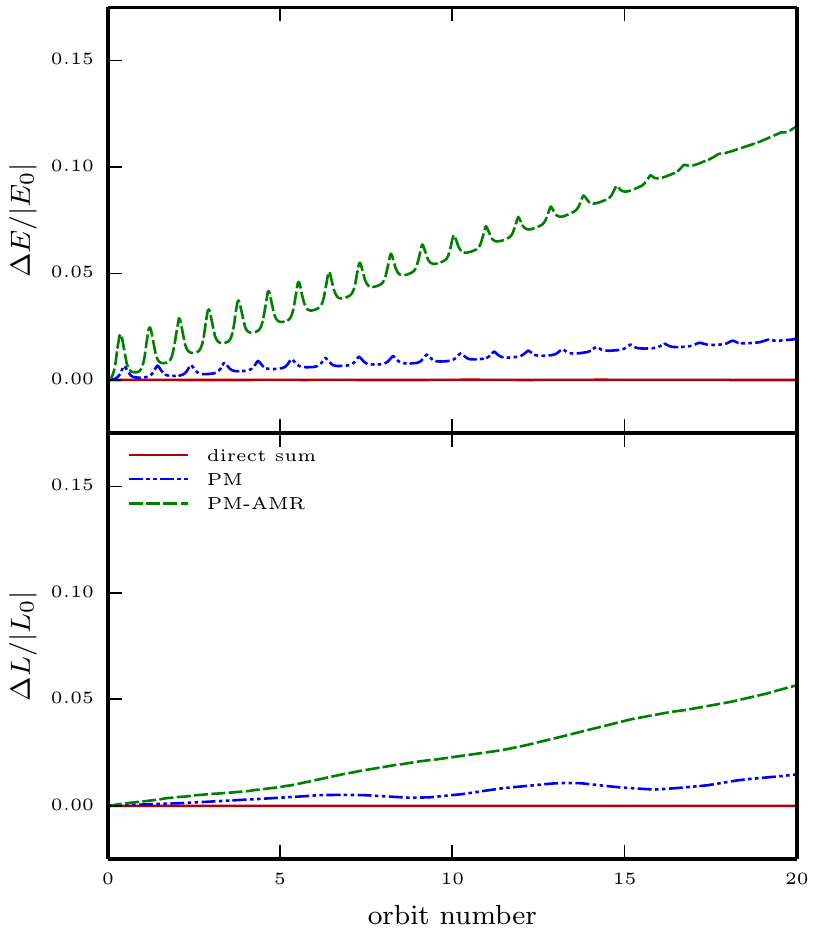}
	\caption{Conservation of energy and angular momentum in the sink binary system obtained using different force calculation methods. The plotted quantities are smoothened over the analytically computed orbital time.}
	\label{fig:orbits}
\end{figure}

The next test concerns the ability of the algorithm to produce accurate sink particle trajectories. We let two equal mass sink particles orbit their common center of mass on elliptic trajectories in the absence of gas. The initial separation of the two sinks is 24 cells and the initial velocities are chosen such that the minimum separation of the two sinks is 6 cells if they move on their analytically predicted orbits. We use this setup to compare the PM force calculation for sink particles with the direct force summation. We furthermore distinguish the PM case into a run where the grid is fully refined to level 7 (PM case) and another setup where the cells further than $6 \Delta x$ from the sink particle are allowed to de-refine to level 6 (PM-AMR case). When AMR is activated we enforce single time stepping, meaning that the coarser level is updated using the same time step as the finer level. The Plummer softening length is set to $2\Delta x$ in the direct force summation run and the radius of the particle `swarm' is set to $3\Delta x$ for the PM cases. We measure total energy and angular momentum in the system during the first 20 orbits and plot the results in Figure \ref{fig:orbits}.
\newline
The direct force summation delivers excellent results in this test. For this setup the time step is controlled by the free fall time criterion (see Equation \ref{freefall}) which leads to a constant time step as long as the sink masses are constant. For a constant time step, the particle integration scheme is equivalent to a leapfrog integrator and therefore obtains its symplectic property. Angular momentum and energy are thus conserved to machine precision. We see a considerable precession of the perihelion by $\approx -7.7^{\circ}$ per orbit which is caused by the deviation from the $1/r$-potential induced by the softening. The results for the PM scheme in the absence of AMR are surprisingly good. The picture changes dramatically when AMR is turned on. The main source of problems for the PM scheme are the level boundaries. The poisson solver in \textsc{ramses} uses a `one-way interface' scheme \citep{Guillet2011} which means that the coarse level potential is used to set boundary conditions for the refined regions. This is problematic since a poorly resolved mass distribution (as it is the case for the sink on coarse levels) leads to large errors in the potential.

%%%%%%%%%%%%%%%%%%%%%%%%
% sink speed
%%%%%%%%%%%%%%%%%%%%%%%%
\subsection*{Appendix B: Sink Integration Speed \label{sink_speed}}
The previous section shows the superior accuracy of the direct force summation over the PM approach. However, there is still good reason to use the PM scheme in order to accelerate calculations involving a `large' number of sinks. In this short subsection we estimate the speed gain that can be expected when using the PM method. We consider an initially homogeneous, slightly turbulent gas sphere at level 8. We then randomly place $n_{\text{sink}}$ equally massive sink particles inside the sphere. The total gas mass is identical to the total sink mass. We then let the code refine around the sink particles up to level 18. As soon as the refinements are done and the usual load balancing has been performed, we measure the time needed to perform ten time steps. Wo obtain the following speedup when using the PM method compared to direct force summation: 
%\begin{table}
%	\caption{Speedup obtained using the PM method instead of direct force summation.}
%	\label{table:pm_speedup}
\begin{center}
	\renewcommand{\arraystretch}{1.2}
	\begin{tabularx}{0.66\columnwidth}{ l r r r r}
		$n_{\text{sink}}$	&10	&100	&1000 	&10000\\
		\midrule															
		PM-acc		&1.1	&1.7		&6.3		 &46.3
	\\														
	\end{tabularx}\\
\end{center}	
%\end{table}
In all cases, the number of gas cells hosted by each MPI process is much larger than the total number of sink particles. When computing the forces directly, it is therefore the $n_{\text{sink}}$ loops over all its cells that each MPI process has to perform which are dominating the extra execution time. We can therefore estimate that a usual hydro and gravity time step by \textsc{ramses} roughly takes takes execution time of 100 loops over all cells. So if $n_{\text{sink}} \gg 100$ the total execution time is dominated by the direct force summation and we recommend switching to the PM method. Here our results differ from what \cite{Federrath2010} find for their implementation into the FLASH code. When computing the direct sum of all sink-gas interactions, their total execution time is not significantly increased for $\sim 1000$ sinks. 

%\section*{Appendix A: Other applications of the clump finder}\label{other_apps}
%We compute additional properties of the clumps. A full list of all clumps and their properties is printed to the log file every time the clump finder is running. Running the clump finder with a low density threshold and a very required peak-to-saddle ratio (choose infinity 

%\subsection{Clumpfinder comparison \label{clumpfinder_test}}
%In this section, we compare the RAMSES clumpfinder to Clumpfind \citep{williams1994}, a tool mainly used by observers. 
%\begin{figure}
%\includegraphics[width=\columnwidth]{/Users/ableuler/zbox/paper_plots/clumps/cmf.pdf}
%\caption{Comparison of the}
%\label{cmf_fig}
%\end{figure}

\end{document}